\documentclass[aps,pra,twocolumn, amssymb,showpacs,superscriptaddress,notitlepage,longbibliography]{revtex4-1}

\usepackage[colorlinks=true,linkcolor=blue,anchorcolor=red,citecolor=blue, urlcolor=blue]{hyperref}
\usepackage{bm}
\usepackage{graphicx}
\usepackage{times}

\begin{document}

\title{Hidden-symmetry-protected $Z_2$ topological insulator in a cubic
 lattice}
\author{Jing-Min Hou}
\email{Corresponding author:jmhou@seu.edu.cn}
\affiliation{School of Physics,
Southeast University, Nanjing  211189, China}

\author{Wei Chen}
\affiliation{College of Science, Nanjing University of Aeronautics and
Astronautics, Nanjing 210016, China}

\begin{abstract}
Usually  $Z_2$ topological insulators are protected by   time reversal symmetry.
Here, we present a new type of $Z_2$ topological  insulators in a cubic lattice which is protected by a novel hidden symmetry, while  time reversal symmetry is broken. The   hidden symmetry  has a composite antiunitary operator consisting of fractional translation, complex conjugation,  sublattice
exchange, and local gauge transformation.     Based on the hidden symmetry, we define the hidden-symmetry polarization and $Z_2$ topological invariant to characterize the topological insulators. The surface states have band structures with odd number of Dirac cones, where pseudospin-momentum locking occurs. When the hidden-symmetry-breaking perturbations   are added on the boundaries, a gap opens in the surface band structure, which confirms that the topological insulator and the surface states are protected by the hidden symmetry. We aslo discuss the  realization and detection of this new kind of $Z_2$ topological insulator in   optical lattices with ultracold atom techniques.
\end{abstract}
\maketitle

 \section{Introduction}
Recently, topological phases in condensed matters attract much attention of physicists\cite{Hasan10rmp,QiXL11rmp}. Before 1980s, it was believed that matter is classified by symmetries according to Laudau's theory. The discovery of  the   quantum Hall effect  overturned that belief since two distinct quantum Hall insulators may have the same symmetry\cite{Klitzing80prl}. Very soon, it was realized that   quantum Hall insulators are classified by a topological invariant, i.e., the Chern number, which is directly related  to the quantized Hall
conductivity\cite{Thouless82prl}. Thus, quantum Hall insulators are time-reversal-symmetry-breaking  topological phases  due to the existence of magnetic field. Since then, the door of the study of topological phases in condensed matter physics was opened. Later, the discoveries of $Z_2$ topological insulators in two and three dimensions significantly boom the research on topological phases in condensed matter physics\cite{Kane05prl_1,Kane05prl_2,Bernevig06prl,Bernevig06sci,Konig07sci,
FuL07prl,Moore07prb,Roy09prb,FuL07prb,Hsieh08nat,XiaY09np,ZhangHJ09np,ChenYL09sci,Hsieh09nat,Hsieh09prl}. In general, the $Z_2$ topological insulators are induced by spin-orbit coupling and  protected by time reversal symmetry. Such nontrivial phases are characterized by the topological edge or surface states, which exhibit  spin-momentum locking.

Besides time-reversal-symmetry-protected topological insulators, there are also topological insulators protected by spatial symmetry (i.e. topological crystalline insulators) that have been predicted theoretically and prepared experimentally\cite{FuL11prl,Hsieh12nc,LiuCX14prb}. Recently, we found a kind of hidden symmetry which protects the degeneracies at Dirac points of a square lattice\cite{HouJM13prl,HouJM14prb}. This kind  of hidden symmetry is a composite antiunitary symmetry, generally  consisting of fractional translation, complex conjugation,  sublattice
exchange, and local gauge transformation. We have found a two-dimensional optical lattice preserving this kind of hidden symmetry, which supports quantum pseudospin Hall effect, i.e., a $Z_2$ two-dimensional topological insulator\cite{HouJM16pra}. A natural question is whether the hidden symmetry supports the existence of  $Z_2$ topological insulators in three dimensions. In this paper, we will give a positive answer.

In the following sections, we propose a tight-binding  model  in a cubic   lattice, which  preserves a hidden symmetry, i.e.,  a composite antiunitary symmetry. Based on the hidden symmetry, the pseudospin, symmetry-polarization and $Z_2$ topological invariant are defined. We calculate the dispersion relation of the system  and find that the band  inversions happen when changing the parameters across some fixed values.  Based on the $Z_2$ topological invariants and band inversions, the phase diagram is drawn.  We evaluate the surface states of a slab geometry and calculate the pseudospin textures  of the surface states.   The insulator with non-trivial topological invariant has odd number of  Dirac cones in the surface band structure, which have pseudospin-momentum-locking pseudospin textures. When the hidden-symmetry-breaking perturbations  are added on the boundaries of the slab, a gap opens at the surface Dirac points and  the surface states on the two opposite boundaries mix, even turn into bulk states when the perturbations are strong enough, which confirm that the topological insulator is protected by the hidden symmetry. The recent development of    experimental techniques  of ultracold atoms in optical lattices have make them become a platform to simulate the exotic physics in condensed matters\cite{Jaksch05anp,Bloch08rmp,Lewenstein07adp}. Thus, we suggest to realize this model with ultracold atoms in optical lattices and to detect the topological properties with state-of-the-art techniques in cold atomic physics.

\begin{figure}[ht]
\includegraphics[width=0.53\columnwidth]{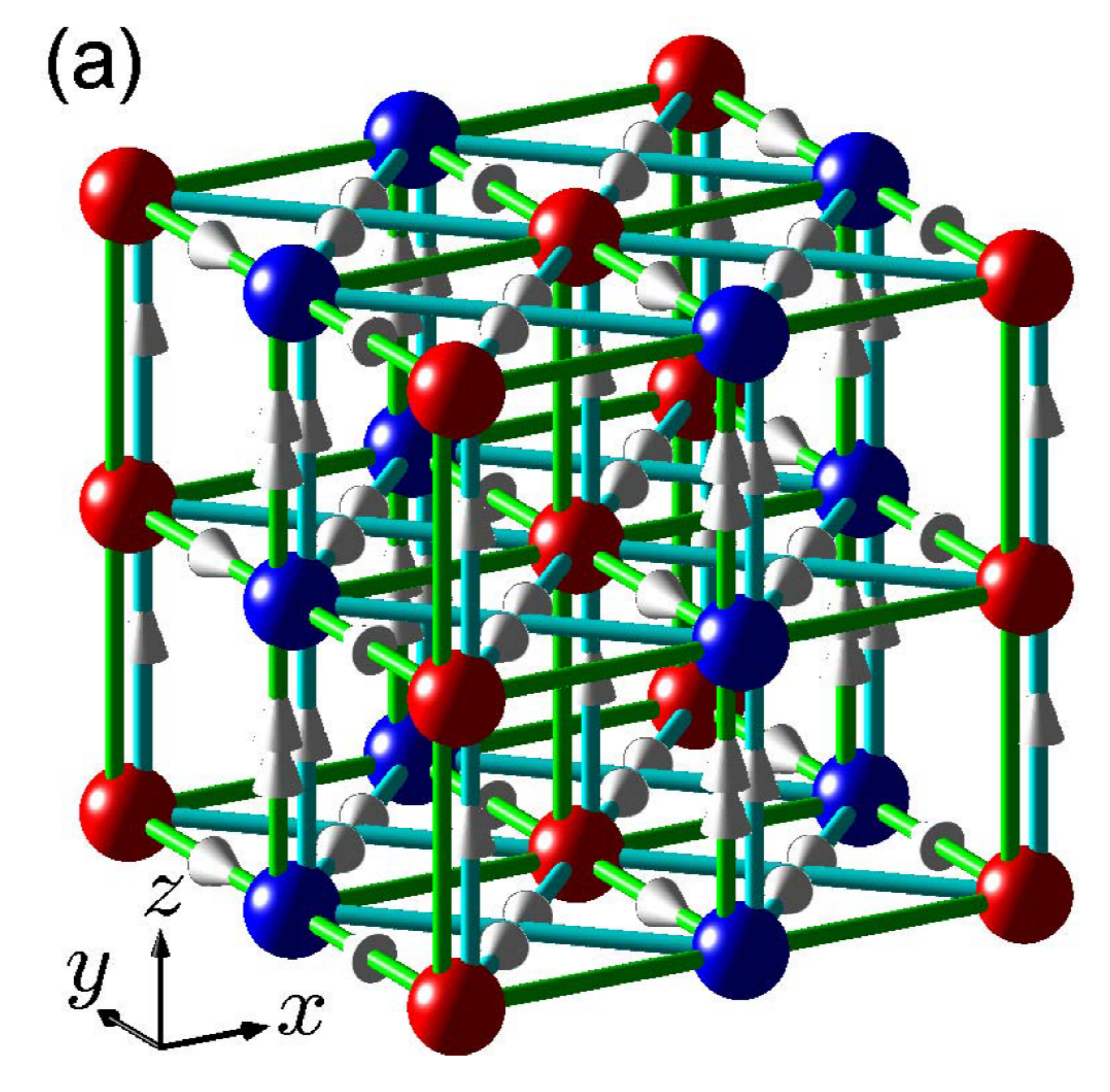}
\includegraphics[width=0.45\columnwidth]{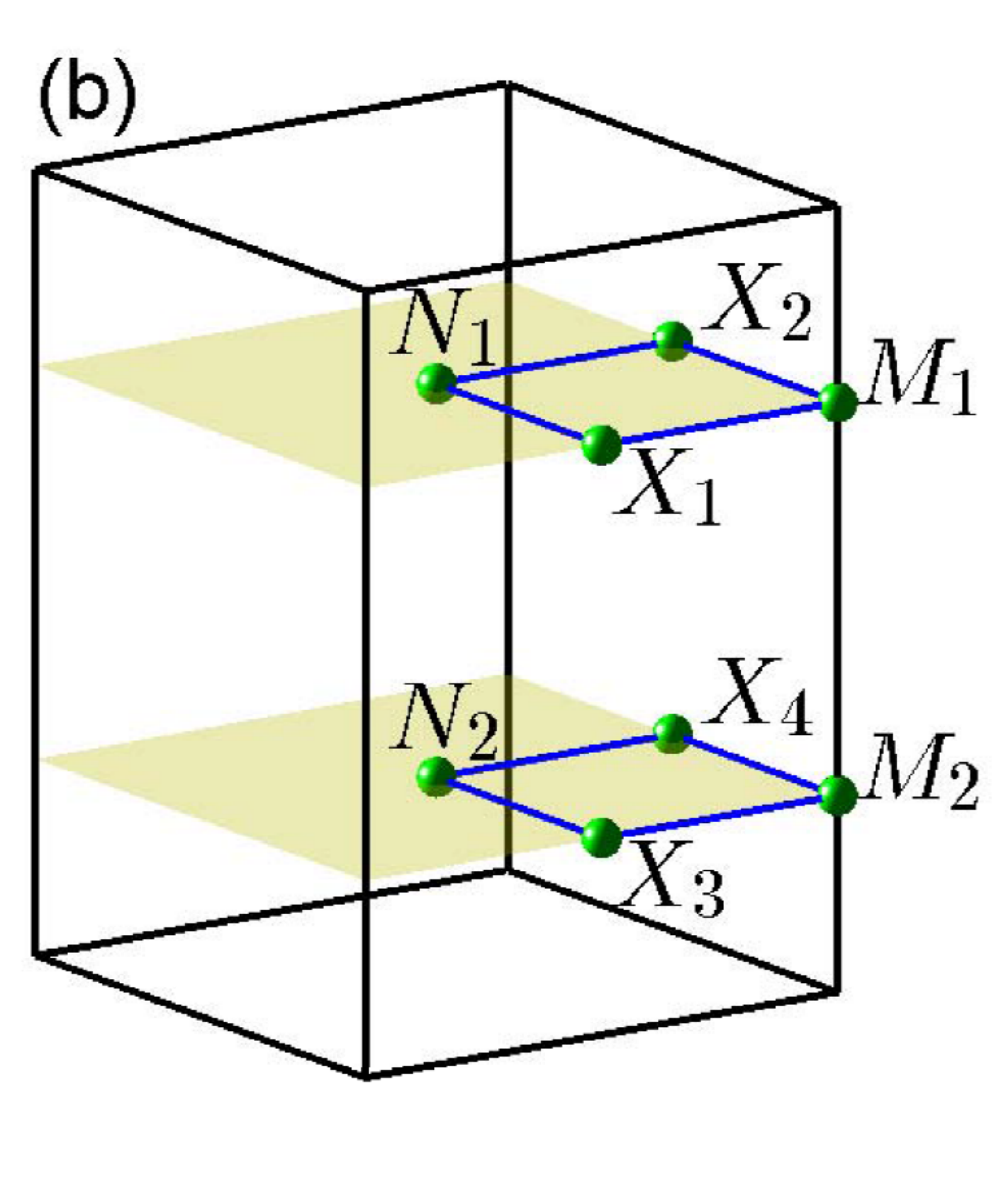}
\caption{  (a)  Schematic of the cubic  lattice, where the red and blue spheres represent the lattice sites of sublattices $A$ and $B$, respectively, and the green and cyan sticks represent the hoppings with the color transition matrices $\tau_x$ and $\tau_z$, respectively, and the  grey single arrows and double arrows denote  the hopping-accompanying phases $\pi/2$ and $\pi$, respectively.    (b) The Brillouin zone, where the green spheres represent the hidden-symmetry-invariant points. }\label{fig1}
\end{figure}

\section{Model}

 Here, we consider two-level atoms trapped in a cubic optical  lattice as shown in Fig.\ref{fig1}(a),
where the arrows represent the hopping-accompanying phases.    Due to the appearing of the
hopping-accompanying phases, the translation symmetry is broken.
Thus, the lattice is divided into two sublattices, i.e. sublattices
$A$ and $B$, denoted by the red and blue spheres in Fig.\ref{fig1}(a), respectively. Taking the distance between the nearest  lattice
sites  as the  unit of length, we define the primitive lattice vectors as
$\mathbf{a}_1=(1,-1,0)$, $\mathbf{a}_2=(1,1,0)$,
 and $\mathbf{a}_3=(0,0,1)$. The primitive reciprocal lattice
 vectors are
$\mathbf{b}_1=(\pi,-\pi,0)$, $\mathbf{b}_2=(\pi,\pi, 0)$, and
$\mathbf{b}_3=(0,0,2\pi)$ and the corresponding Brillouin zone is shown in Fig.\ref{fig1}(b).
The system can be described by the tight-binding
Hamiltonian  $H=H_0+H_1+H_2$ with
 \begin{eqnarray}
  H_0&=&-\sum_{i\in A} [t_x\hat a^\dag_{i}\tau_x\hat
b_{i+\hat{x}}+t_x\hat a^\dag_{i}\tau_x\hat
b_{i-\hat{x}}\nonumber\\
&&+t_ye^{-i\pi/2} \hat a^\dag_{i}\tau_x\hat b_{i+\hat{y}}+t_ye^{-i\pi/2} \hat a^\dag_{i}\tau_x\hat b_{i-\hat{y}}\nonumber\\
&&+t_z\hat a^\dag_{i}\tau_x\hat a_{i+\hat{z}}-t_z \hat b^\dag_{i+\hat{x}}\tau_x\hat b_{i+\hat{x}+\hat{z}}]\nonumber\\
&&+H.c.  \label{tbh1}
\end{eqnarray}
and
\begin{eqnarray}
  H_1&=&-t_1\sum_{i\in A} [a^{\dag}_{i}\tau_z
a_{i+\hat{x}-\hat{y} }- a^{\dag}_{i}\tau_z
a_{i+\hat{x}+\hat{y}}   +e^{-i\pi/2}a^{\dag}_{i} \tau_z  a_{i+ \hat{z}} ] \nonumber\\
&& -t_1\sum_{i\in B} [b^{\dag}_{i}\tau_z
b_{i+\hat{x}-\hat{y} }- b^{\dag}_{i}\tau_z
b_{i+\hat{x}+\hat{y} }     +e^{-i\pi/2}b^{\dag}_{i} \tau_z  b_{i+ \hat{z}} ]\nonumber\\
&&+H.c.\label{tbh2}
\end{eqnarray}
and
\begin{eqnarray}
H_2=\lambda\sum_{i\in A}a_i^\dag \tau_z a_i+\lambda\sum_{i\in B}b_i^\dag\tau_z b_i\label{tbh3}
\end{eqnarray}
where $a_i=[a_i^{(1)}, a_i^{(2)} ]^T$ and
$b_i=[b_i^{(1)},b_i^{(2)}]^T$ are the two-component  annihilation
operators destructing an atom at a lattice site of sublattice $A$
and $B$, respectively; $\tau_i (i=x,y,z)$   represent the Pauli
matrices in the color space spanned by the two atomic levels; $t$ and $t_1$ represent the amplitudes
of hopping between the nearest lattice sites and between the
next-nearest lattice sites, respectively; $\lambda$ is the magnitude of an effective Zeeman term.

\begin{figure}[ht]
\includegraphics[width=0.45\columnwidth]{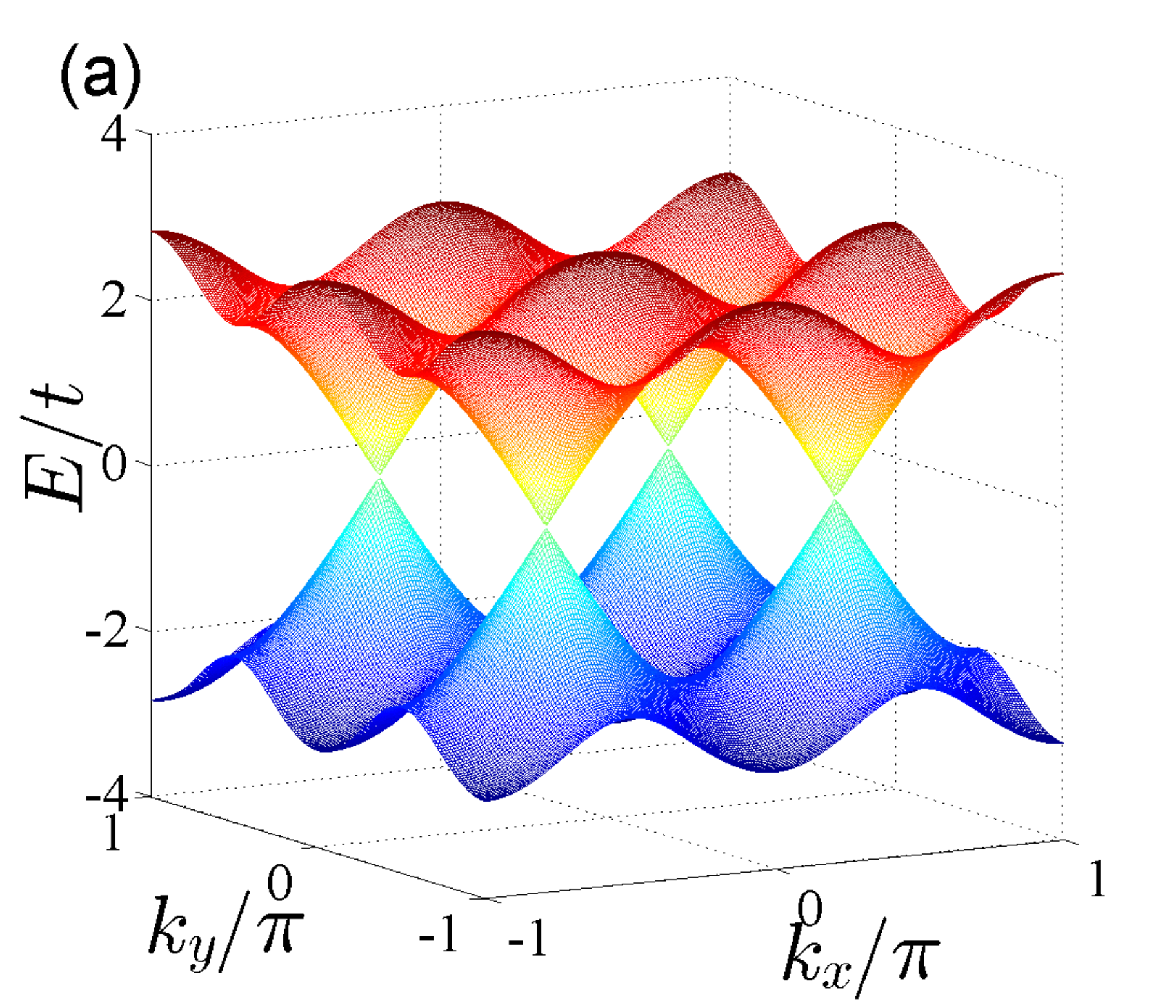}
\includegraphics[width=0.45\columnwidth]{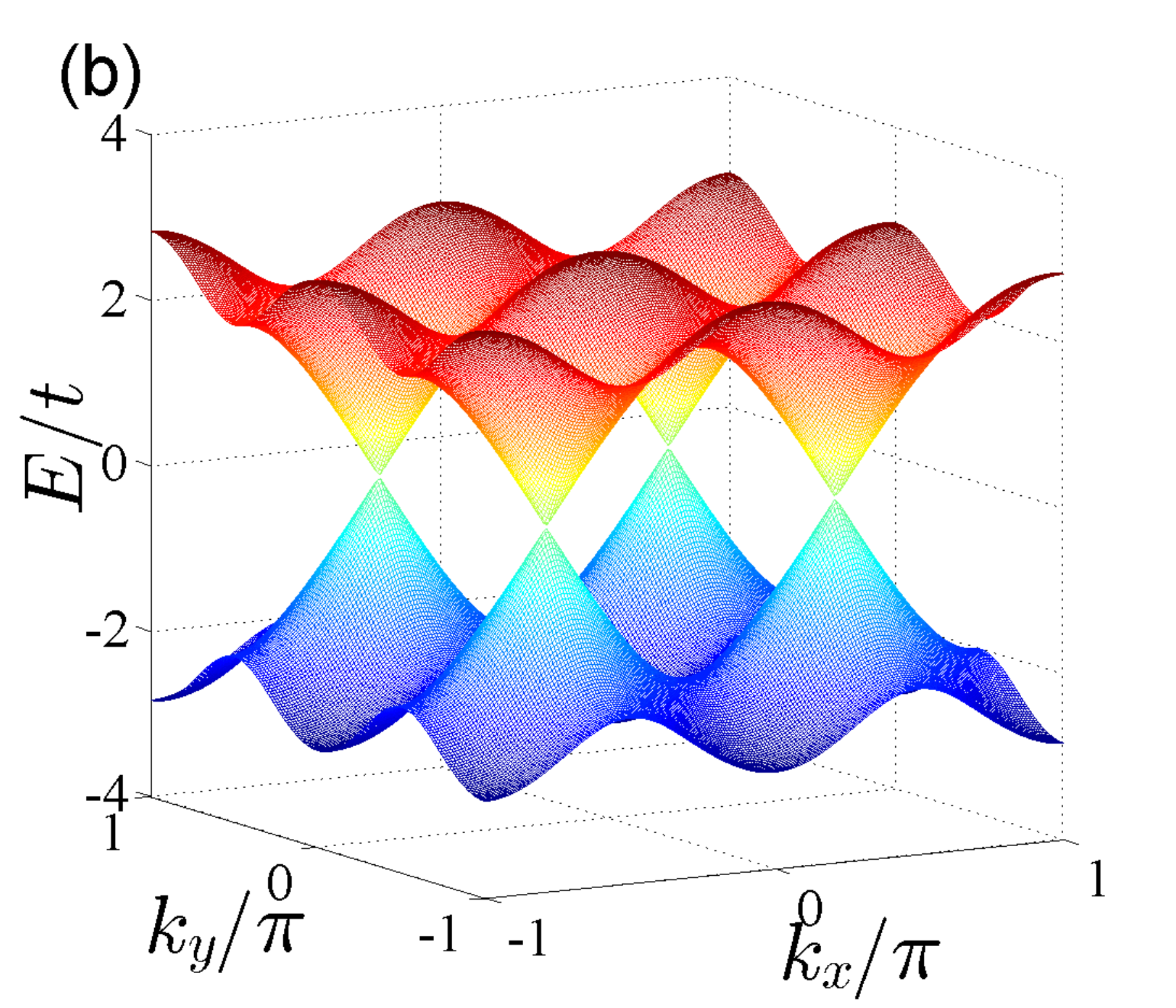}
\includegraphics[width=0.45\columnwidth]{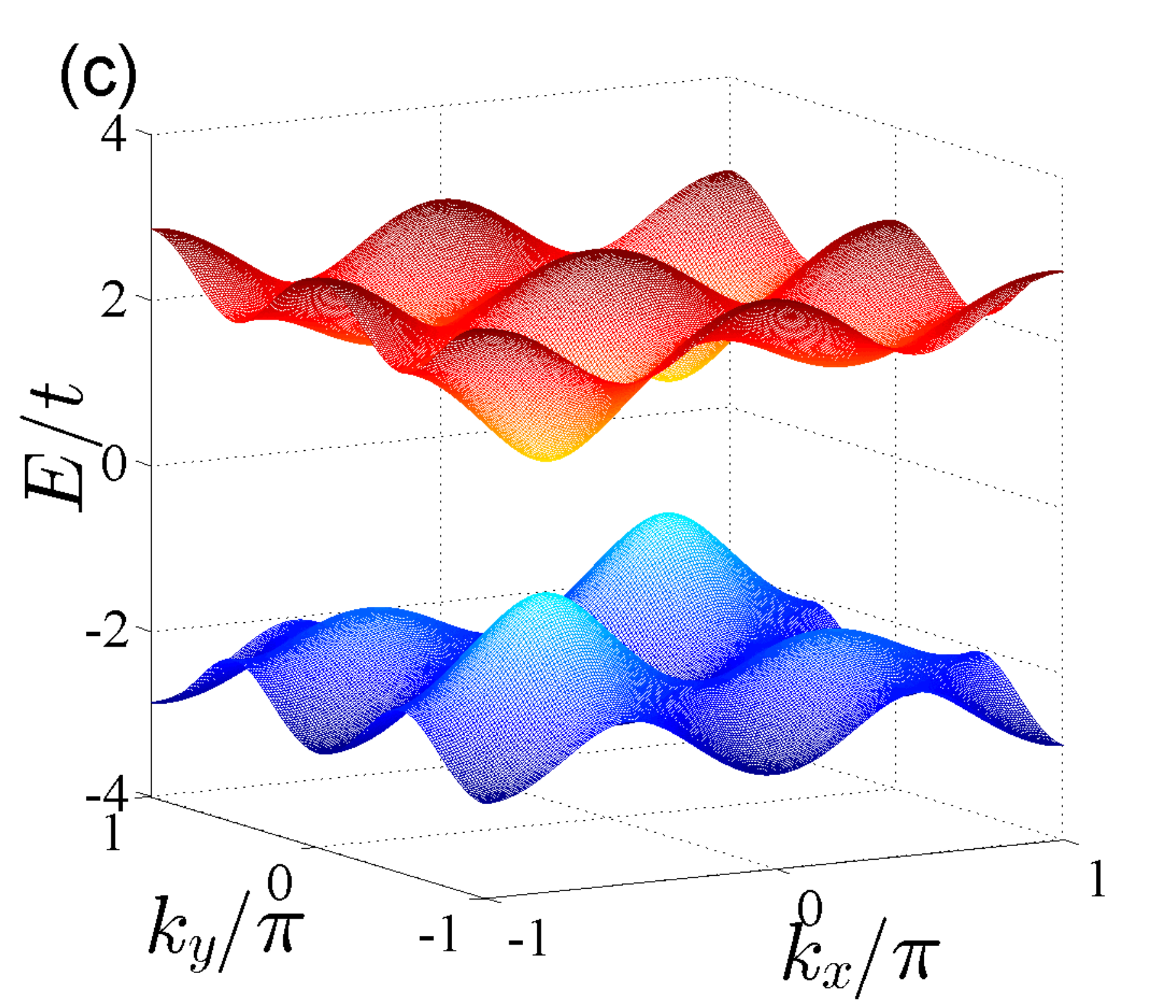}
\includegraphics[width=0.45\columnwidth]{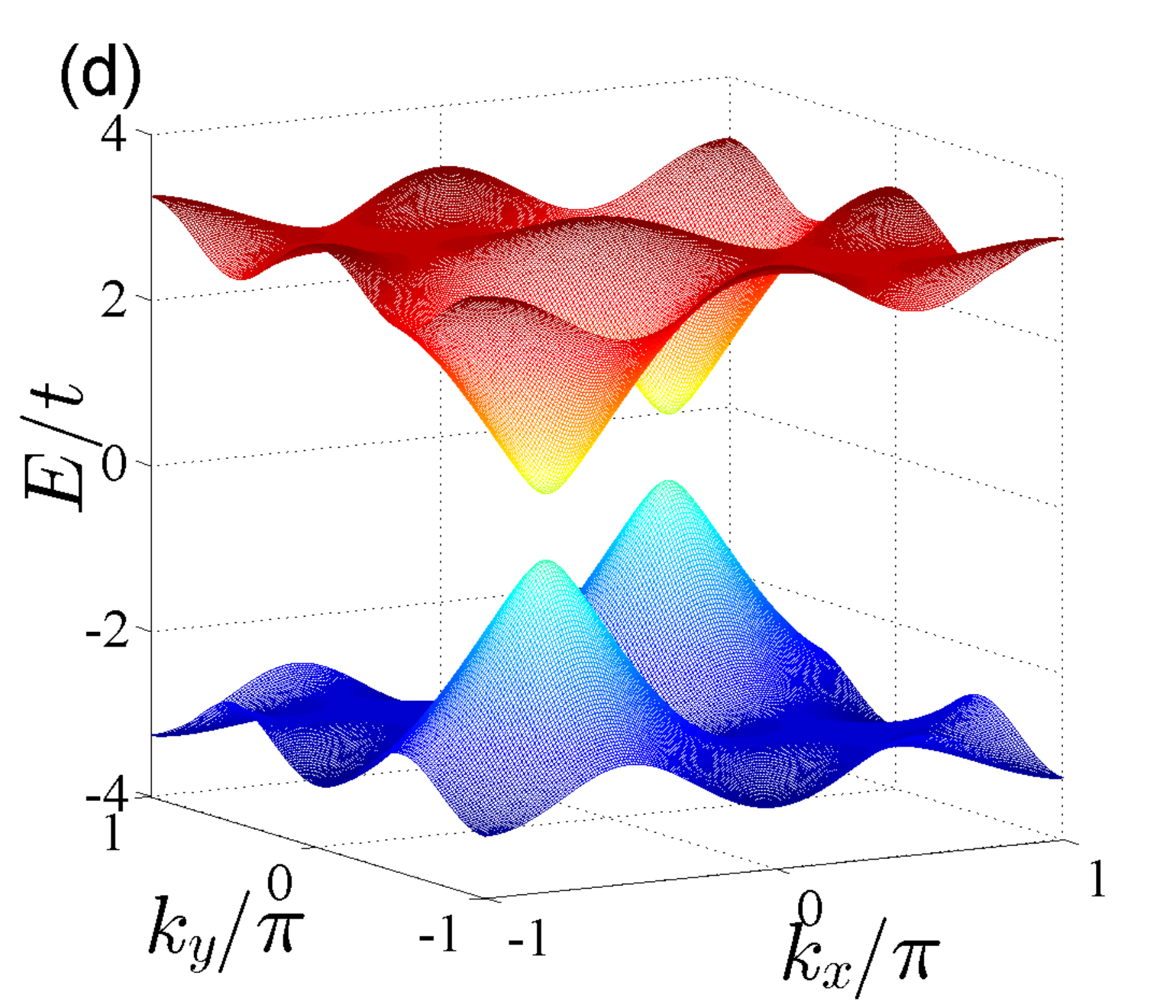}
\includegraphics[width=0.45\columnwidth]{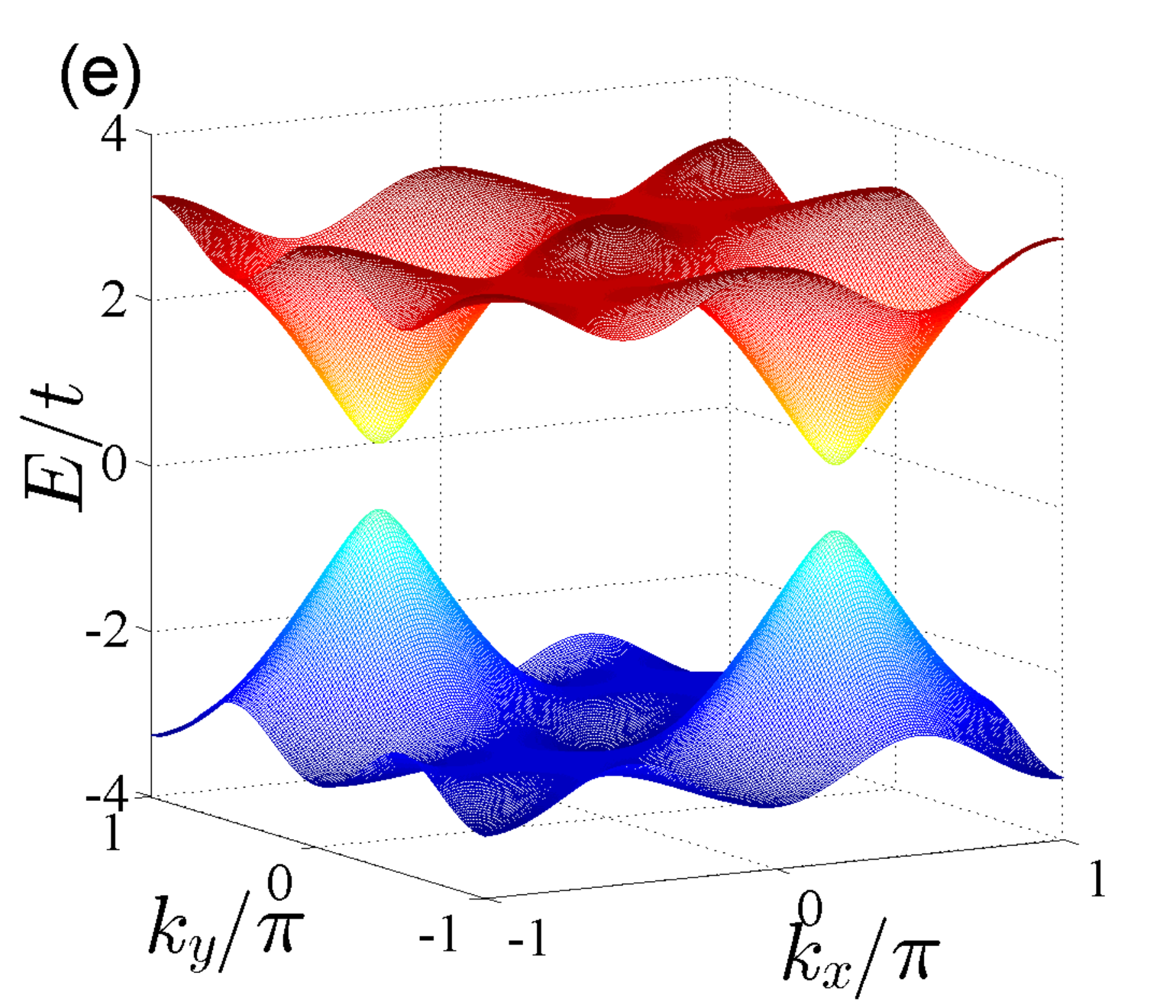}
\includegraphics[width=0.45\columnwidth]{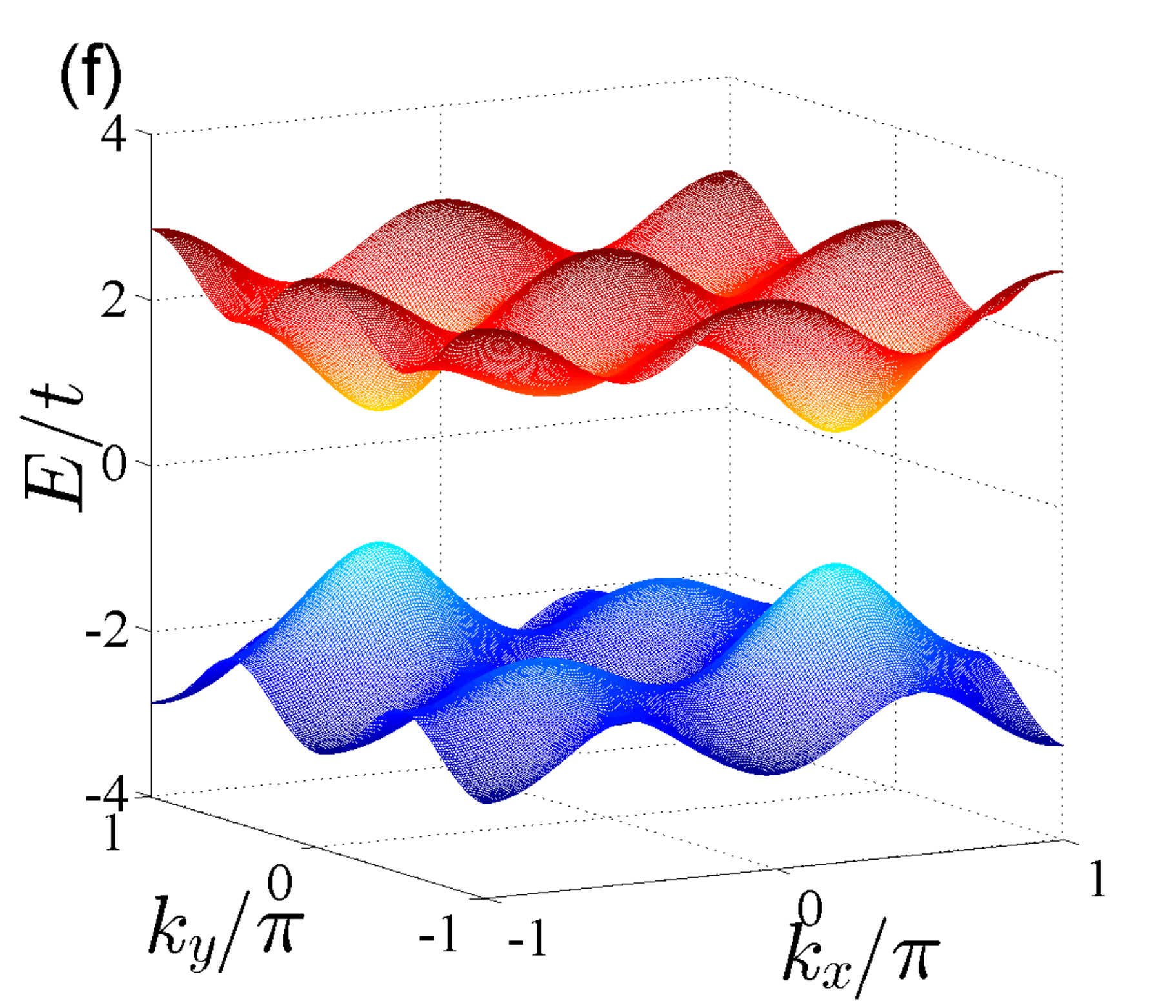}
\caption{  The dispersion relations  (a) on the $k_z=\pi/2$ plane and  (b) on the $k_z=-\pi/2$ plane for $t_1=0$ and $\lambda=0$,   (c) on the $k_z=\pi/2$ plane and  (d) on the $k_z=-\pi/2$ plane for $t_1=0.3t$ and $\lambda=t$,  (e) on the $k_z=\pi/2$ plane and  (f) on the $k_z=-\pi/2$ plane for  $t_1=0.3t$ and $\lambda=-t$. }\label{band}
\end{figure}

\begin{figure}[ht]
\includegraphics[width=0.9\columnwidth]{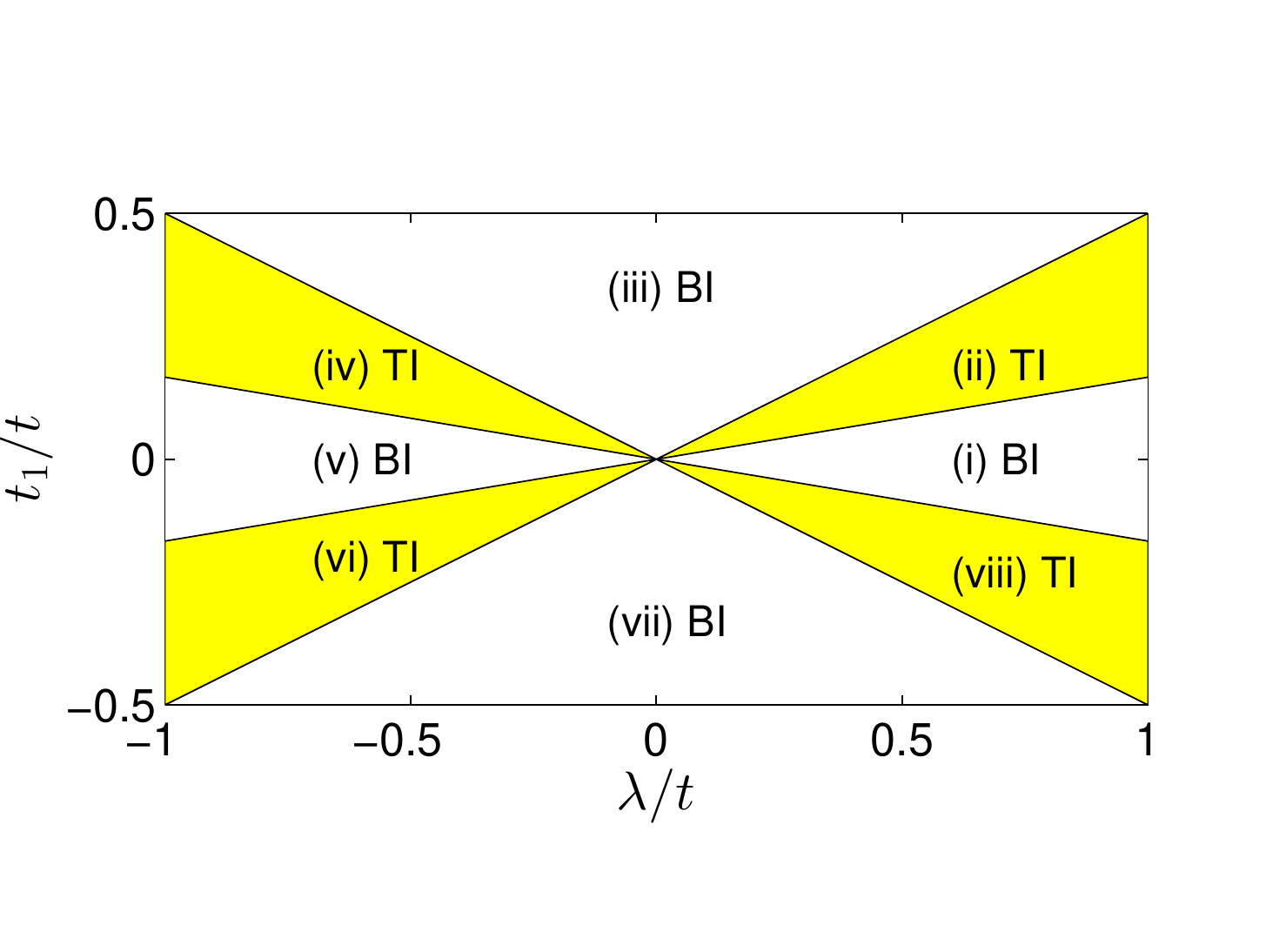}
\caption{    The phase diagram of the cubic lattice. Here, TI and BI represent topological insulators and conventional band insulators, respectively. }\label{phasedia}
\end{figure}

After  the Fourier transformation, the Bloch Hamiltonian is obtained  as
\begin{eqnarray}
{\cal H}(\mathbf{k})&=&-2t\cos k_x \sigma_x\otimes\tau_x-2t\cos k_y \sigma_y\otimes\tau_x
\nonumber\\
  &&-2t\cos k_z \sigma_z\otimes\tau_x
  +m(\mathbf{k}) I\otimes \tau_z,\label{BH}
\end{eqnarray}
where $m(\mathbf{k})=\lambda -4t_1\sin k_x \sin k_y -2t_1\sin k_z$ represents a mass term. Diagonalizing Eq.(\ref{BH}), we arrive at  the dispersion relation as
\begin{eqnarray}
E(\mathbf{k})=\pm\sqrt{4t^2(\cos^2k_x+ \cos^2k_y+ \cos^2k_z)+m(\mathbf{k})^2}
\end{eqnarray}
From the dispersion relation, it is found that the energy bands are two-fold degenerate for the conduction and valence bands.
When   Hamiltonians $H_1$ and $H_2$ disappear, the mass term vanishes and the conduction and valence bands touched at the points $X_{1,2}=(\pi/2,\mp\pi/2, \pi/2)$ and $X_{3,4}=(\pi/2,\mp\pi/2, -\pi/2)$  in the Brillouin zone as shown in Figs.\ref{band}(a) and (b). When  $H_1$ and $H_2$ appear, the mass term is non-zoro, then a gap opens between the conduction and valence bands as shown in Figs.\ref{band}(c)-(d). The masses at the Dirac points have different signs for different parameter  ranges as shown in Table \ref{tab1}.  Based on the signs of masses at the Dirac points, they can be divided  into eight parameter ranges: (i) $\lambda>6|t_1|$, (ii)
$t_1>0$ and  $6t_1>\lambda> 2t_1$, (iii)
$|\lambda|<2t_1$, (iv)
$t_1>0$ and $-2t_1>\lambda>-6t_1$, (v)
$\lambda<-6|t_1|$, (vi)
$t_1<0$ and $ 2t_1>\lambda>6t_1$, (vii)
$|\lambda|<-2t_1$,  (viii)
$t_1<0$ and  $-6t_1>\lambda>-2t_1$. These insulators are classified by a $Z_2$ topological invariant, which will be defined in Section \ref{topo}. As one crosses the boundary between two neighboring parameter ranges, a band inversion happens at one of Dirac points, which indicates   a topological phase transition. That is to say, the insulators in    two neighboring parameter ranges belong to two distinct topological sectors. The insulators with odd number of negative masses at the Dirac points are   topologically non-trivial insulators, while the ones with even number of negative masses at the Dirac points are trivial insulators, which will be confirmed in the succeeding sections of the paper. Therefore, the system is a $Z_2$ topological insulator in parameter ranges (ii) (iv), (vi) and (viii) and is a trivial band insulator in parameter ranges (i), (iii), (v) and (vii) as shown in Fig.\ref{phasedia}.

\begin{table}
\begin{tabular}{|c|c|c|c|c|c|}
\hline
&Parameter ranges&$X_1$&$X_2$&$X_3$&$X_4$\\
\hline
(i)&$\lambda>6|t_1|$&$+$&$+$&$+$&$+$\\
\hline
(ii)&$t_1>0,  6t_1>\lambda> 2t_1$&$+$&$-$&$+$&$+$\\
\hline
(iii)&$|\lambda|<2t_1$&$+$&$-$&$+$&$-$ \\
\hline
(vi)&$t_1>0, -2t_1>\lambda>-6t_1$&$-$&$-$&$+$&$-$\\
\hline
(v)&$\lambda<-6|t_1|$&$-$&$-$&$-$&$-$\\
\hline
(vi)&$t_1<0, 2t_1>\lambda>6t_1$&$-$&$+$&$-$&$-$\\
\hline
(vii)&$|\lambda|<-2t_1$&$-$&$+$&$-$&$+$ \\
\hline
(viii)&$t_1<0,  -6t_1>\lambda>-2t_1$&$+$&$+$&$-$&$+$\\
\hline
\end{tabular}
\caption{The sign of the masses at the Dirac points for different parameter ranges.}
\label{tab1}
\end{table}

\section{Hidden symmetry}
It is easy to verify that the system has a hidden symmetry with the symmetry operator as
\begin{eqnarray}
\Upsilon=(e^{i\pi})^{i_z}(\sigma_x\otimes I)T_{\hat{x}}K
\label{hs}
\end{eqnarray}
where $K$ is the complex conjugate
operator, $T_{\hat{x}}$ is a translation operator that moves
the lattice by a unit along the $x$ direction, $\sigma_x$ is the Pauli matrix representing sublattice
exchange, $I$ is the unit matrix in the color space, and $(e^{i\pi})^{i_z}$ is a local $U(1)$ gauge transformation and $i_z$ is the $z$-component of the space coordinate.
The Bloch functions are supposed to have the form
$\Psi_{\mathbf{k}}(\mathbf{r})=[u_{A,\mathbf{k}}^{(1)}(\mathbf{r}),
u_{A,\mathbf{k}}^{(2)}(\mathbf{r}),u_{B,\mathbf{k}}^{(1)}(\mathbf{r}),
u_{B,\mathbf{k}}^{(2)}(\mathbf{r})]^T
e^{i\mathbf{k}\cdot\mathbf{r}}$ in the coordinate representation. The
symmetry operator $\Upsilon$ acts on the Bloch function as follows
\begin{eqnarray}
\Upsilon\Psi_{\mathbf{k}}(\mathbf{r})&=&\left(\matrix{u^{(1)*}_{B,\mathbf{k}}(\mathbf{r}-\hat{x})e^{ik_x}\cr
u^{(2)*}_{B,\mathbf{k}}(\mathbf{r}-\hat{x})e^{ik_x}\cr
u^{(1)*}_{A,\mathbf{k}}(\mathbf{r}-\hat{x})e^{ik_x}\cr
u^{(2)*}_{A,\mathbf{k}}(\mathbf{r}-\hat{x})e^{ik_x}}\right)e^{-i\mathbf{k}\cdot\mathbf{r}+i\pi i_z}
=\Psi'_{\mathbf{k}'}(\mathbf{r}).\label{Bloch2}
\end{eqnarray}
 Because $\Upsilon$  is the symmetry operator of the system,
$\Psi'_{\mathbf{k}'}(\mathbf{r})$ must be a Bloch   function of
the system. Thus, we obtain $\mathbf{k}'=(-k_x, -k_y, -k_z+\pi)$,
$u^{(n)}_{A,\mathbf{k}'}(\mathbf{r})=u^{(n)*}_{B,\mathbf{k}}(\mathbf{r}-\hat{x})e^{ik_x}$
and
$u^{(n)}_{B,\mathbf{k}'}(\mathbf{r})=u^{(n)*}_{A,\mathbf{k}}(\mathbf{r}-\hat{x})e^{ik_x}$
with $n=1,2$.
Thus, one can find that the symmetry operator $\Upsilon$ acts on the wave vector as
\begin{eqnarray*}
\Upsilon:\mathbf{k}=(k_x, k_y, k_z)\rightarrow \mathbf{k}'=(-k_x, -k_y, -k_z+\pi)
\end{eqnarray*}
If $\mathbf{k}' = \mathbf{k} + \mathbf{G}$, where $\mathbf{G}$ is a reciprocal
lattice vector, then we can say that $\mathbf{k}$ is a  $\Upsilon$-invariant point in
momentum space.  There are eight distinct
$\Upsilon$-invariant points as  $X_{1,2,3,4}=(\pi/2, \pm \pi/2,\pm \pi/2)$,   $N_{1,2}=(0,0, \pm \pi/2)$, and $M_{1,2}=(0,\pi,\pm \pi/2)$ in the Brillouin zone as shown in Fig.\ref{fig1}(b). From Eq.(\ref{hs}), we obtain that the square of the hidden symmetry operator is $\Upsilon^2=T_{2\hat{x}}$, which has the representation based on the Bloch
functions as $\Upsilon^2=e^{-i2\mathbf{k}\cdot\hat{x}}$. Therefore, we have $\Upsilon^2=-1$ at points $X_{1,2,3,4}$ while $\Upsilon^2=1$ at points $N_{1,2}$ and $M_{1,2}$ in the Brillouin zone. Furthermore, since $\Upsilon$ is an antiunitary operator,   there must exist two-fold   degeneracies at points $X_{1,2,3,4}$, which are protected by the hidden symmetry $\Upsilon$\cite{HouJM13prl}.

\section{ $Z_2$ topological invariant}
\label{topo}
\begin{figure}[ht]
\includegraphics[width=0.45\columnwidth]{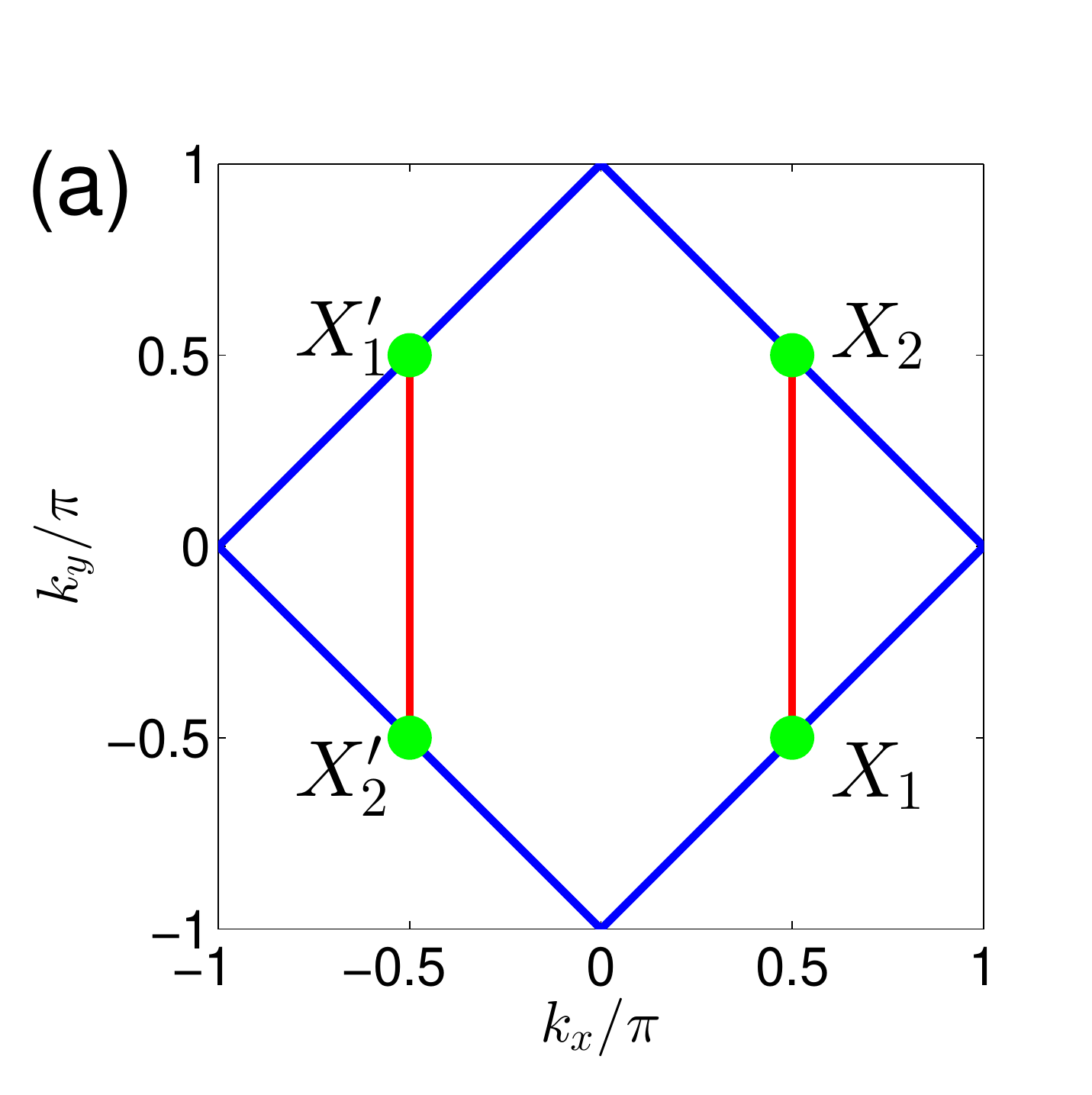}
\includegraphics[width=0.45\columnwidth]{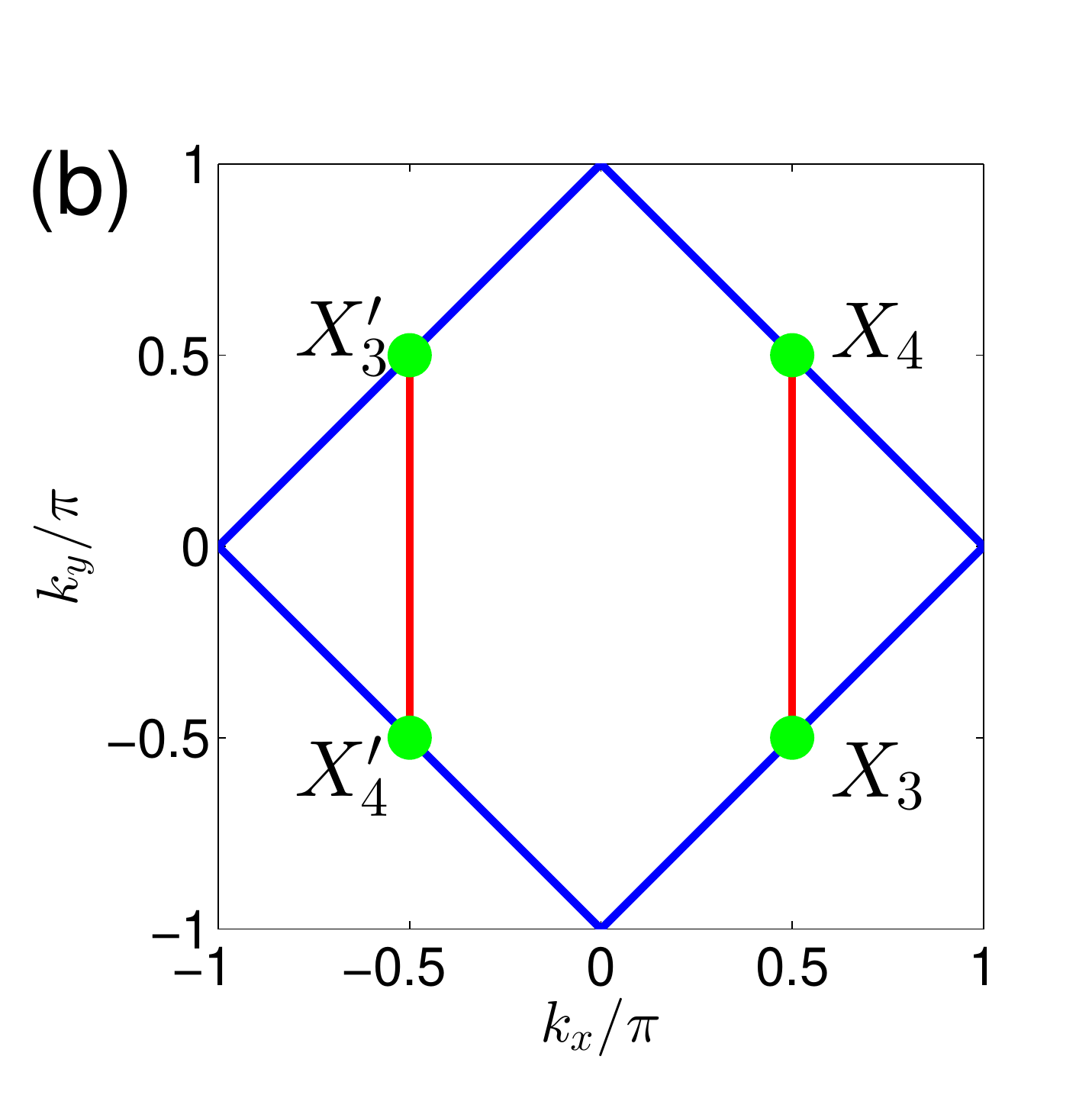}
\caption{ The integration paths for the definition of   charge polarization on (a) the $k_z=\pi/2$ and  (b) $k_z=-\pi/2$ planes in  the Brillouin zone. Here, the red lines represent the integration paths and $X_i$ and $X_i'$ represent the $\Upsilon$-invariant points  at which hidden-symmetry-protected degeneracies appear.    }\label{Brillouin2d}
\end{figure}

Based on the hidden symmetry $\Upsilon$, we can define a $Z_2$ topological invariant, which is used to classify the insulator phases of  the system.
The Bloch  functions of the occupied bands can be written as
 $|\Psi_{n,\mathbf{k}}\rangle=e^{i\mathbf{k}\cdot\mathbf{r}}|u_{n,\mathbf{k}}\rangle$,
 where $|u_{n,\mathbf{k}}\rangle$ is the the cell-periodic eigenstate
 of the Bloch Hamiltonian $\mathcal{H}(\mathbf{k})$.
 The Berry connection matrix is defined as
 \begin{eqnarray}
 \mathbf{a}_{mn}=-i\langle
 u_{m,\mathbf{k}}|\nabla_\mathbf{k}|u_{n,\mathbf{k}}\rangle
 \end{eqnarray}
 For the hidden symmetry $\Upsilon$, we also define a matrix as
\begin{eqnarray}
w_{mn}(\mathbf{k})=\langle
u_{m, \mathbf{k}'}|\Upsilon|u_{n,\mathbf{k}}\rangle\label{wmatrix}
\end{eqnarray}
where $\mathbf{k}'$ is the wave vector by $\Upsilon$ acting on $\mathbf{k}$, i.e., $\mathbf{k}'=(-k_x,-k_y,-k_z+\pi)$.
Since $\Upsilon^2=-1$ is satisfied at $\Upsilon$-invariant points $X_{1,2,3,4}$ and $\Upsilon$ is an antiunitary operator, it is easy to verify    that $w$ is an antisymmetric matrix at the $\Upsilon$-invariant degenerate points $X_{1,2,3,4}$.

For the present model with half filling, there are two occupied bands, which compose the $\Upsilon$ pair bands. For the occupied $\Upsilon$ pair bands,
we   define the charge polarization in terms of the Berry connection as
\begin{eqnarray}
P^{ \pm }_\rho&=&\frac{1}{2\pi}\left[\int_{X_\alpha}^{X_\beta}\mathbf{A}(\mathbf{k})\cdot
d\mathbf{k}+\int_{X_\alpha'}^{X_\beta'}\mathbf{A}(\mathbf{k})\cdot
d\mathbf{k} \right]
\end{eqnarray}
where the integration is along the red lines on the $k_z=\pi/2$ and $k_z=-\pi/2$ planes of the Brillouin zone  as shown in Fig.\ref{Brillouin2d}; $\mathbf{A}(\mathbf{k})$ is defined as
$\mathrm{tr}[\mathbf{a}(\mathbf{k})]$; The signs $ \pm $ denote the $k_z=\pi/2$ and $k_z=-\pi/2$ planes, respectively; $\alpha=1, \beta=2$ for the $k_z=\pi/2$ plane and $\alpha=3, \beta=4$ for the $k_z=-\pi/2$ plane. For each occupied band,   the partial
charge polarization is defined as
\begin{eqnarray}
P^{ \pm }_i&=&\frac{1}{2\pi}\left[\int_{X_\alpha}^{X_\beta}\mathbf{a}_{ii}(\mathbf{k})\cdot
d\mathbf{k}+\int_{X_\alpha'}^{X_\beta'}\mathbf{a}_{ii}(\mathbf{k})\cdot
d\mathbf{k} \right]
\end{eqnarray}
For the $\Upsilon$ pair bands,
we can also define the $\Upsilon$ polarization as
\begin{eqnarray}
P^{ \pm }_\Upsilon&=&P^{ \pm }_1-P^{ \pm }_2=2P^{ \pm }_1-P^{ \pm }_\rho\nonumber\\
&=&\frac{1}{2\pi}\int_{X_\alpha}^{X_\beta}[\mathbf{A}(\mathbf{k})-\mathbf{A}(\mathbf{k}')]\cdot
d\mathbf{k} -\frac{i}{\pi}\log\frac{w_{12}(X_\beta) }{w_{12}(X_\alpha) }\nonumber\\
&=&\frac{1}{i\pi}\log\left[\frac{\sqrt{w_{12}(X_\alpha)^2}}{w_{12}(X_\alpha)}\cdot\frac{w_{12}(X_\beta)}{\sqrt{w_{12}(X_\beta)^2}} \right]
\label{UpsilonPor}
\end{eqnarray}
The Hilbert space
can be classified into two groups depending on the difference
 between the $\Upsilon$ polarizations on the  $k_z=\pi/2$ and $k_z=-\pi/2$ planes,
\begin{eqnarray}
\Delta=P_\Upsilon^{ + }-P^{ - }_\Upsilon\label{Delta}
\end{eqnarray}
The $\Upsilon$ polarization is an interger and   only defined modulo
$2$ due to the  ambiguity of the log. The argument of the log has only two values $\pm 1$ associated with the even and odd values of $P^\pm_\Upsilon$, respectively.
Therefore, we can rewrite Eq.(\ref{Delta}) as
\begin{eqnarray}
(-1)^{\Delta}
=\prod_{\alpha=1}^4\frac{\textrm{Pf}[w(X_\alpha)]}{\sqrt{\textrm{det}[w(X_\alpha)]}}
\label{Z2}
\end{eqnarray}
The $Z_2$ topological invariant can be defined as $\Delta$ modulo $2$. When $\Delta$ is odd or even, the system is a topological insulator or a trivial band insulator.
Therefore, Eq.(\ref{Z2}) gives a distinct definition of the $Z_2$ topological invariant.

\section{Surface states and their pseudospin textures}
\begin{figure}[ht]
\includegraphics[width=0.45\columnwidth]{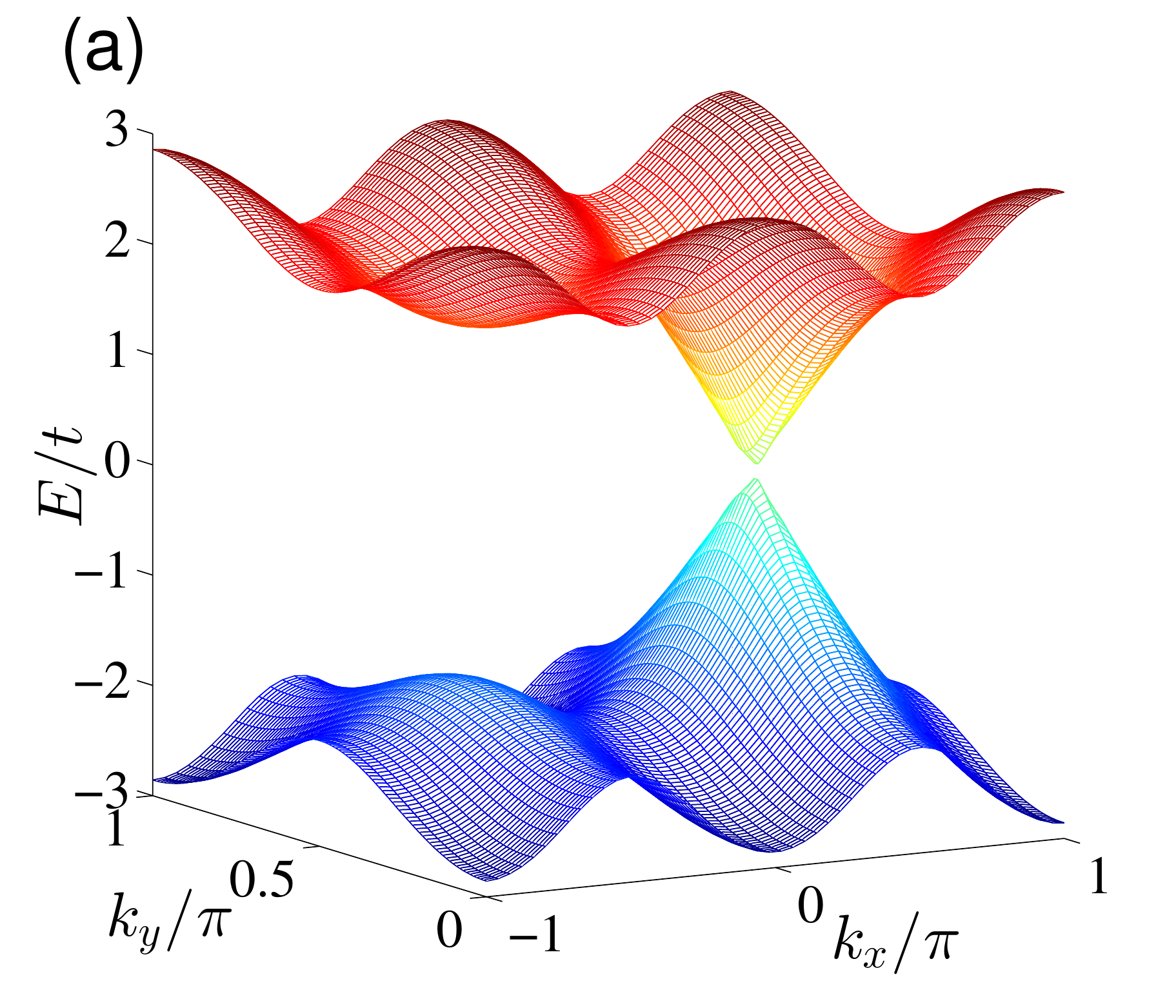}
\includegraphics[width=0.45\columnwidth]{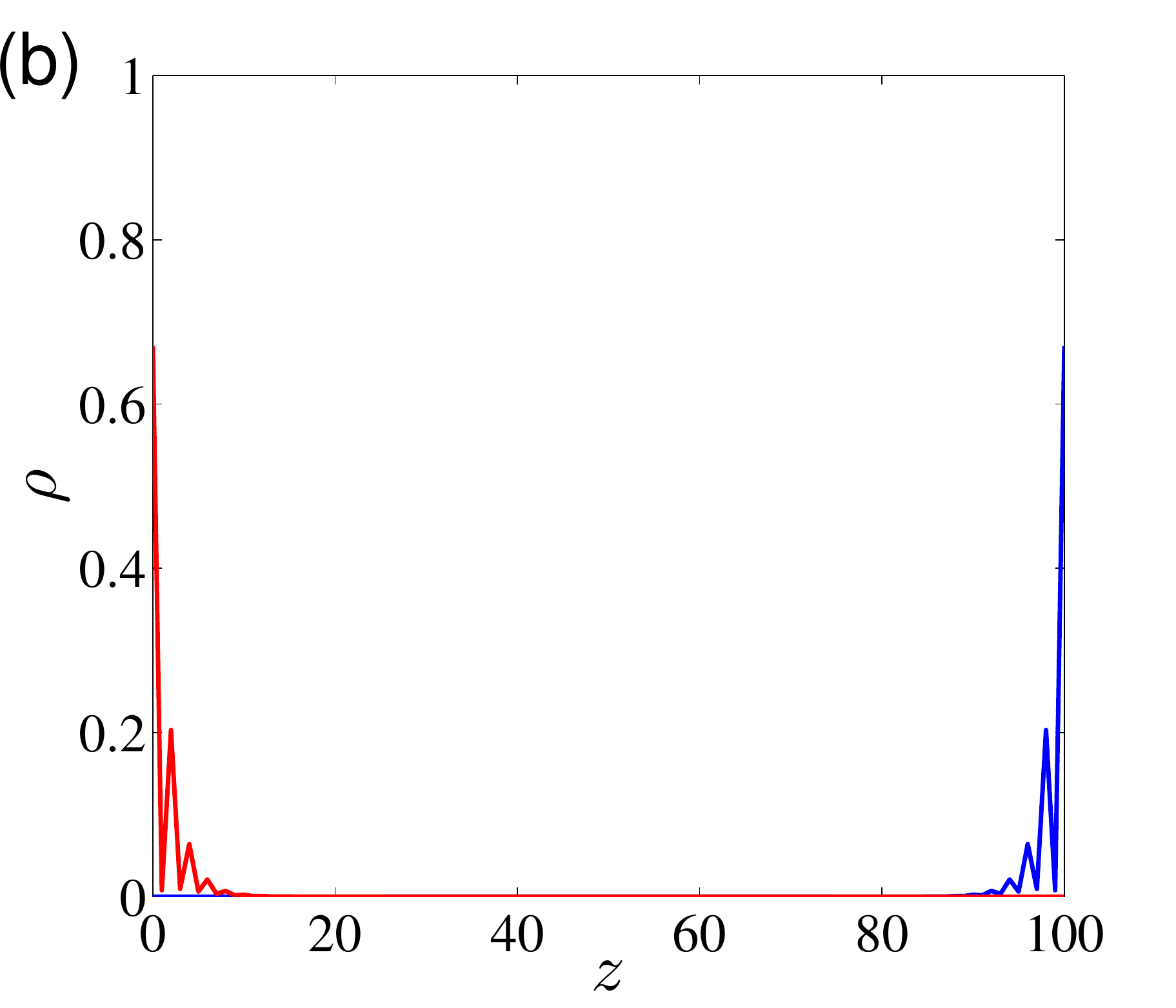}
\includegraphics[width=0.45\columnwidth]{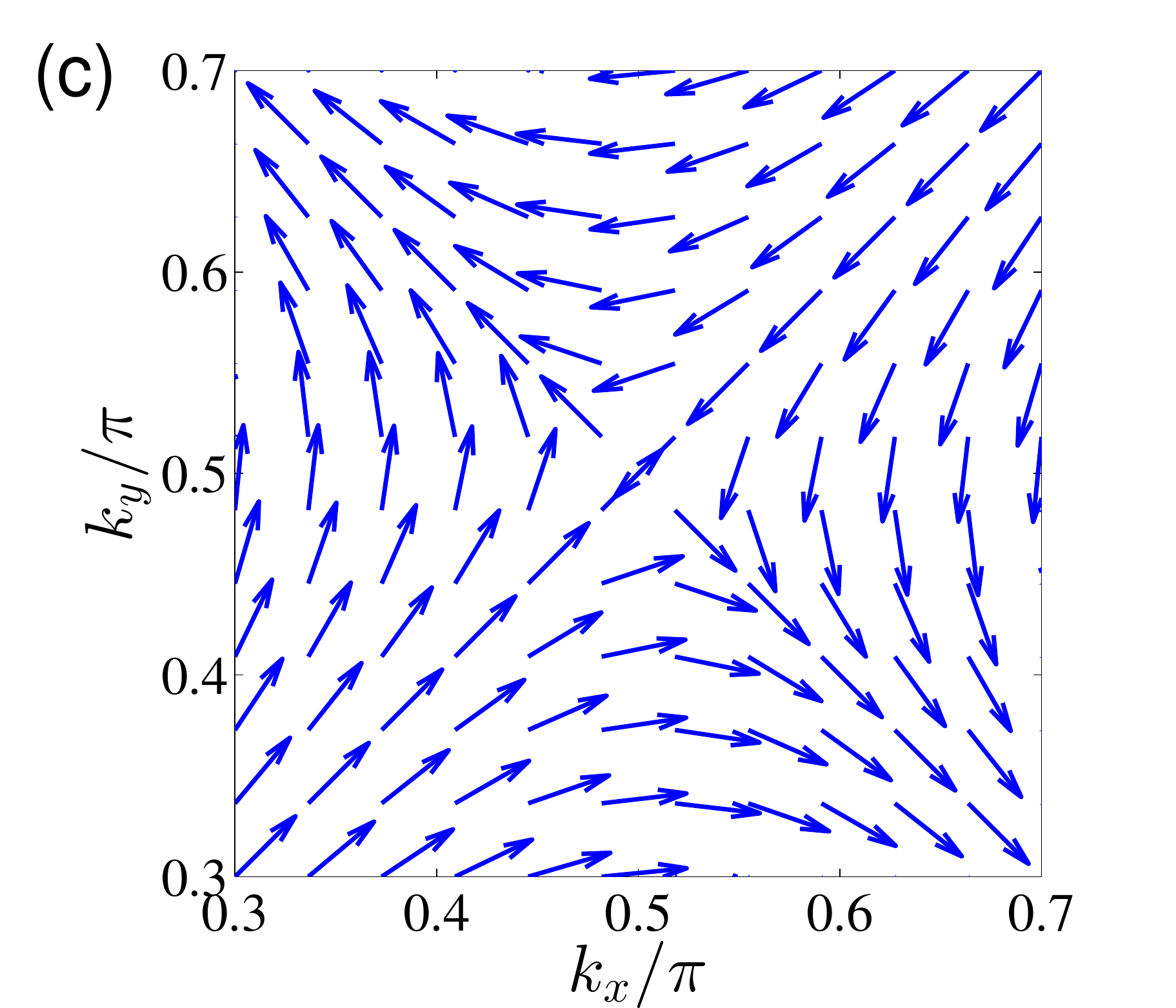}
\includegraphics[width=0.45\columnwidth]{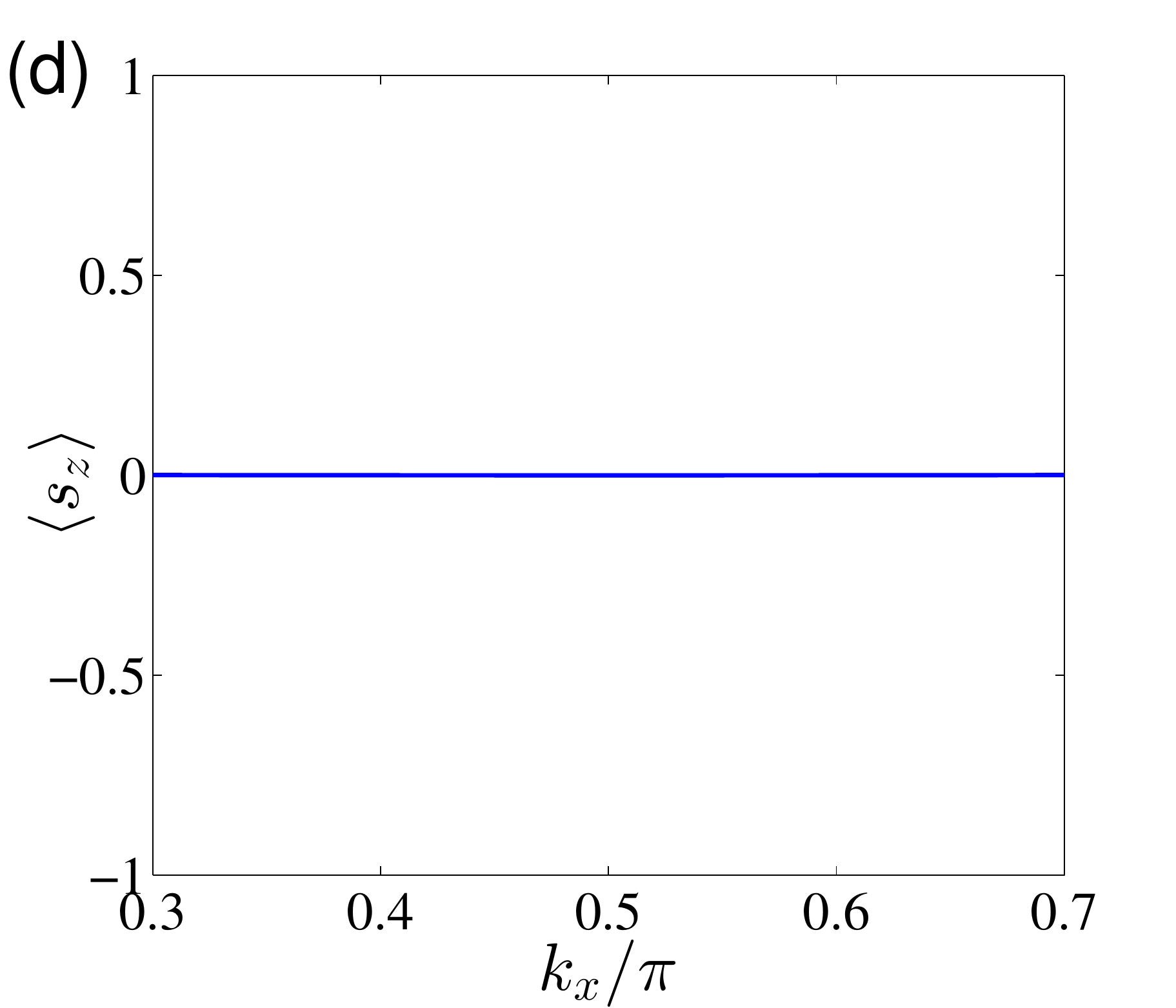}
\caption{ The properties of the highest valence bands (surface states)  of a slab preserving the hidden symmetry for the case $t_1=0.3t$ and $\lambda=t$. (a) The dispersion relation;  (b)   the probability density of the surface states with the wave vector $(k_x,k_y)=(0.6\pi,0.5\pi)$, where the red and blue lines correspond to the two degenerate states localized on the opposite open boundaries, respectively; (c) the pseudospin textures of the average pseudospin components $(\langle s_x\rangle, \langle s_y\rangle)$ on the $k_x$-$k_y$ plane; (d) the profiles of the average  component of pseudospin $\langle s_z\rangle $ along the $k_y=0.5\pi$ line.  Here, the two degenerate    surface states localized on  the two boundaries have the same pseudospin textures and the vanishing average pseudospin component $\langle s_z\rangle $. }\label{surface1}
\end{figure}

\begin{figure}[ht]
\includegraphics[width=0.45\columnwidth]{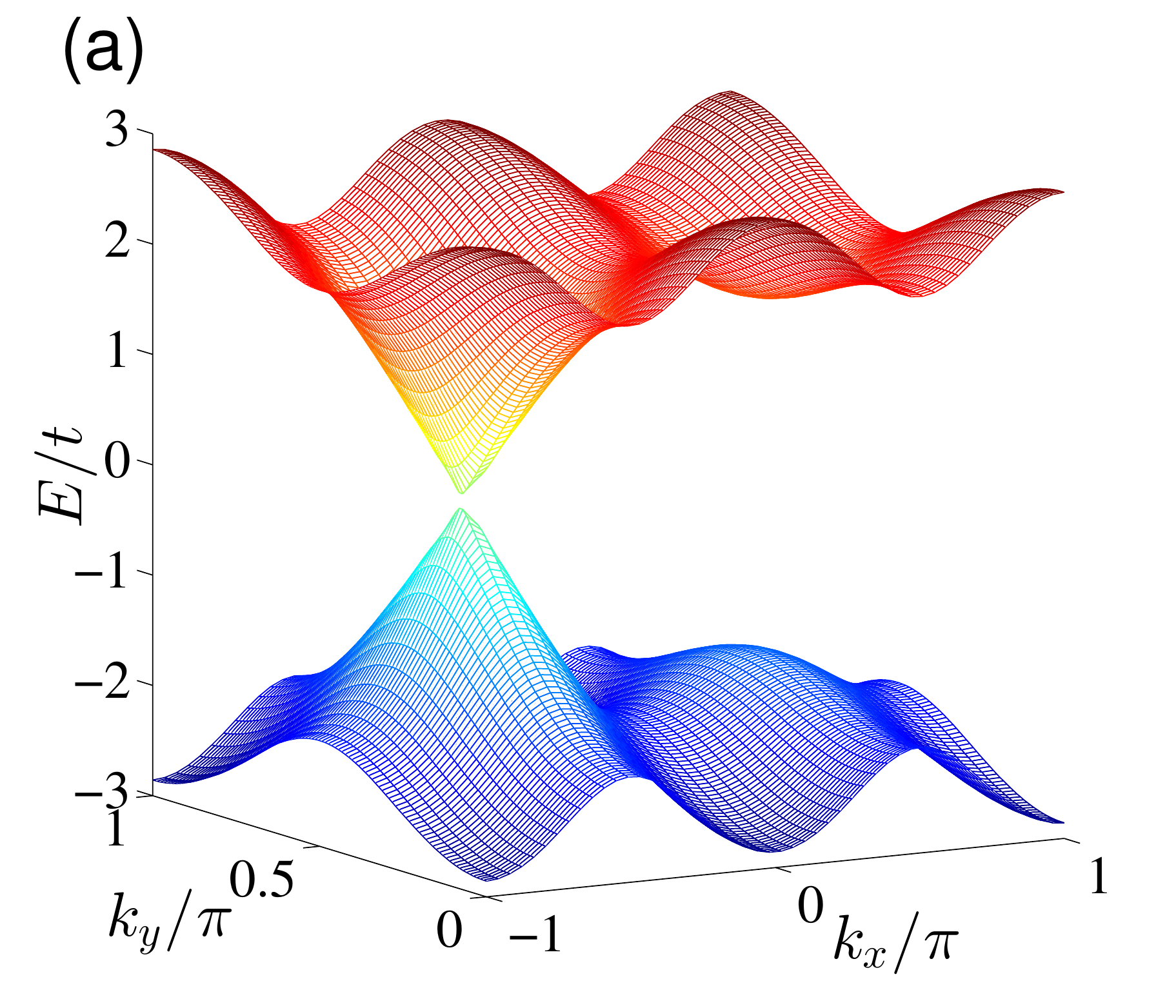}
\includegraphics[width=0.45\columnwidth]{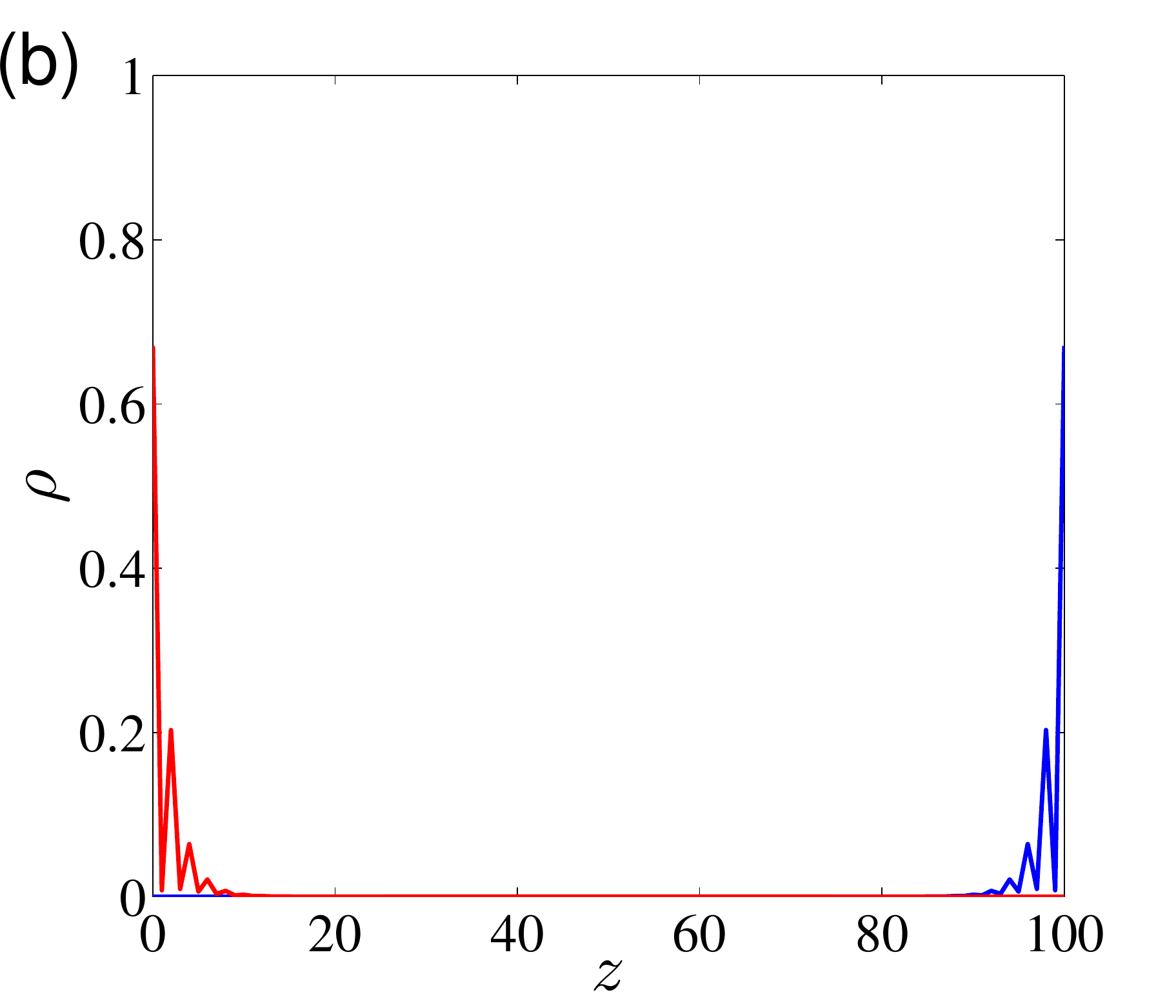}
\includegraphics[width=0.45\columnwidth]{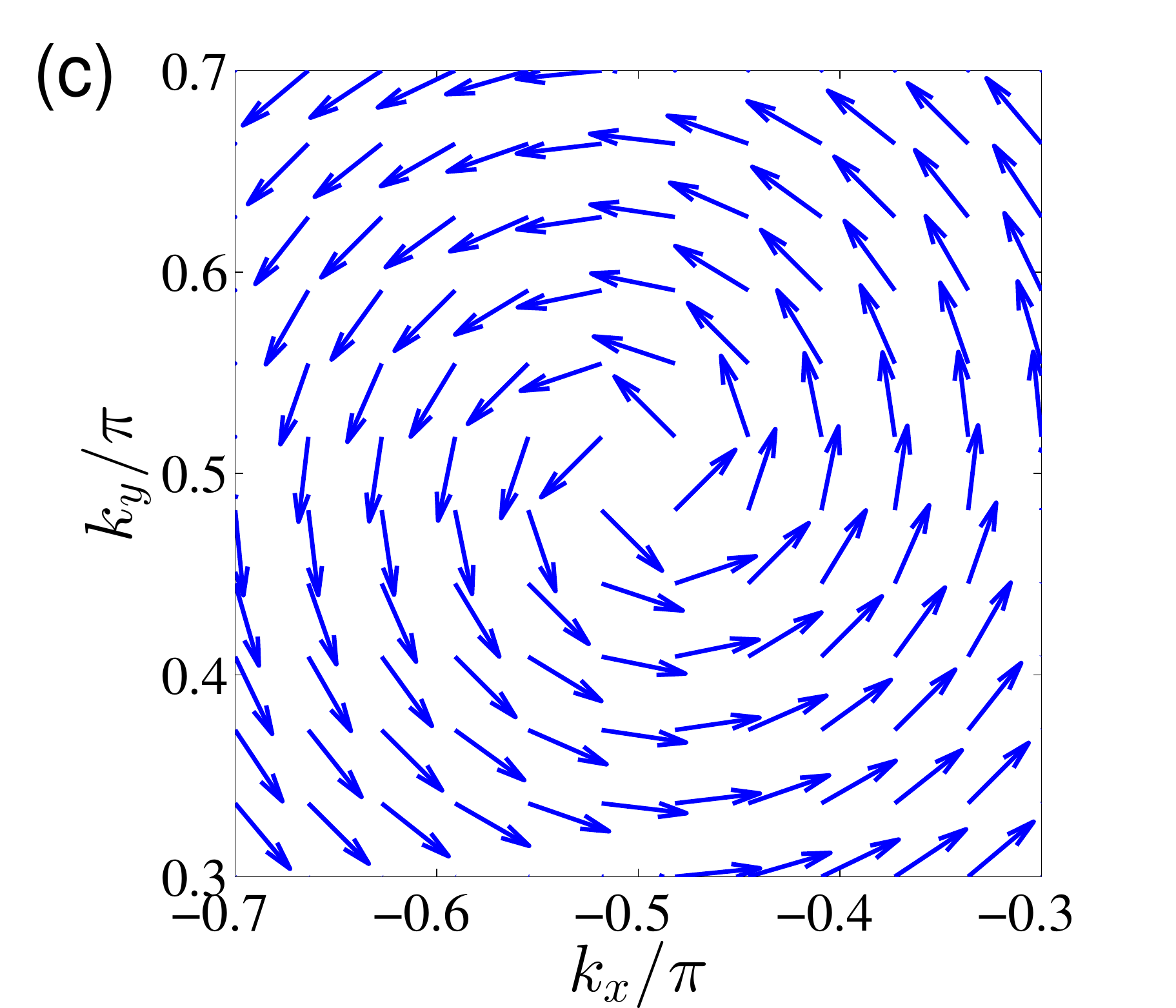}
\includegraphics[width=0.45\columnwidth]{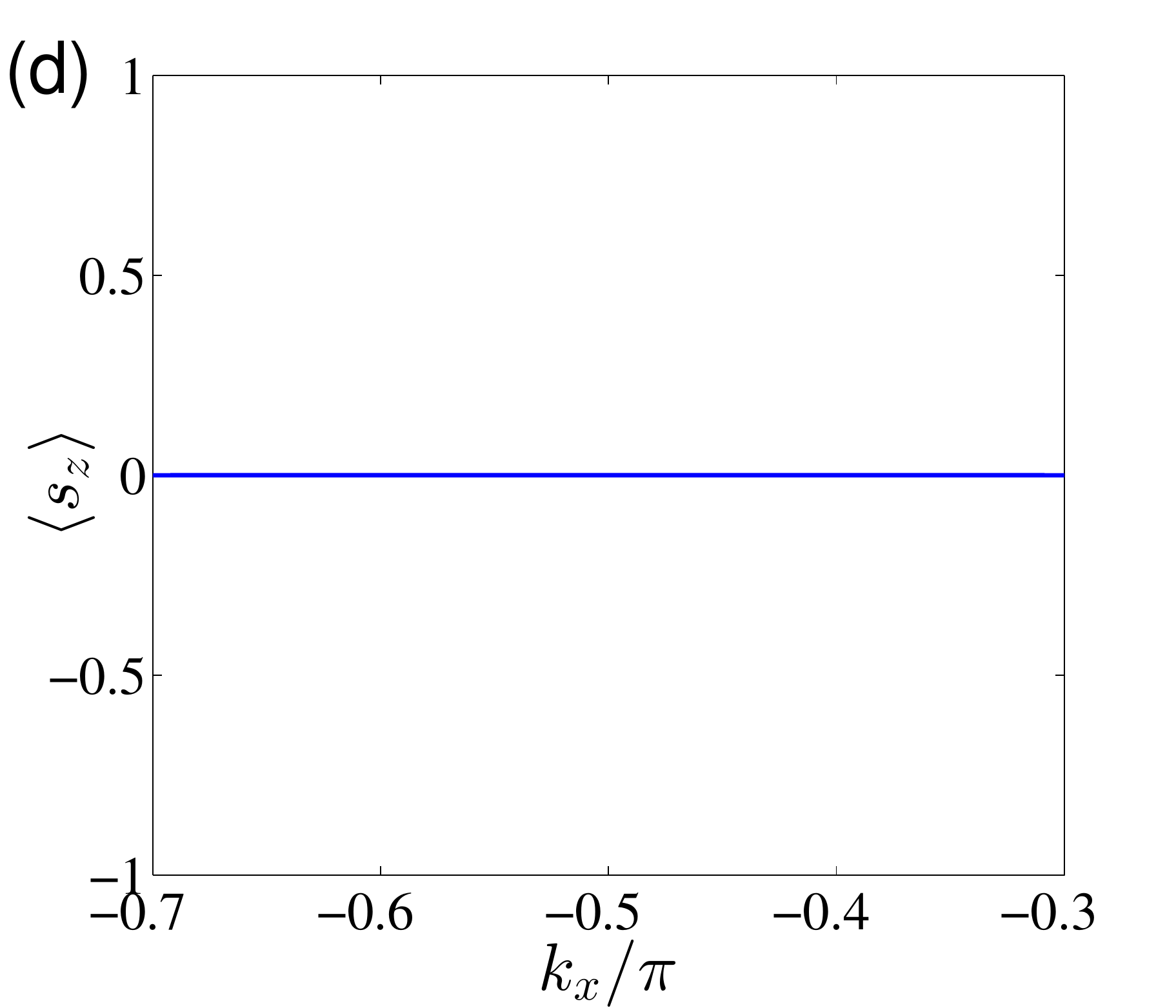}
\caption{ The properties of the highest valence bands (surface states)  of a slab preserving the hidden symmetry for the case $t_1=0.3t$ and $\lambda=-t$.   (a) The dispersion relation;  (b)   the probability density of the surface states   with the wave vector $(k_x,k_y)=(-0.6\pi,0.5\pi)$, where the red and blue lines correspond to the two degenerate states localized on the opposite open boundaries, respectively; (c) the pseudospin texture of the average pseudospin components $(\langle s_x\rangle, \langle s_y\rangle)$ on the $k_x$-$k_y$ plane; (d) the profile of the average  component of pseudospin $\langle s_z\rangle $ along the $k_y=0.5\pi$ line.  Here, the two degenerate    surface states localized on  the two boundaries have the same pseudospin textures and the vanishing average pseudospin component $\langle s_z\rangle $. }\label{surface2}
\end{figure}

Generally,   topological insulators have special surface states, for example,  the surface bands have odd number of Dirac cones in the corresponding surface Brillouin zone. Here, in order to manifest the surface states of the hidden-symmetry-protected $Z_2$ topological insulator, we investigate a slab with open boundaries along the $z$ direction. Based the numerical results, it is found that there exists a single Dirac cone on the surface Brillouin zone for the   topological insulator phases, which    as shown in Fig.\ref{surface1} and Fig.\ref{surface2} for parameter ranges (ii) and (iv), respectively. In the these two parameter ranges, the system is a topological insulator but  the location of  the surface Dirac point in the surface Brillouin zone  are different. In  parameter range (ii), the surface Dirac point  locates   at points $(\pi/2, \pi/2)$  of the surface Brillouin zone as shown in  Fig.\ref{surface1}(a), while,  in  parameter range (iv), the surface Dirac point  locates   at point  $(-\pi/2, \pi/2)$ of the surface Brillouin zone as shown   Fig.\ref{surface2}(a).  For every state on the surface Dirac cone,    the distribution  of probability density concentrate on one of the open boundaries as shown in Fig.\ref{surface1}(b) and Fig.\ref{surface2}(b). The Dirac cone surface states  are two-fold degenerate, since they localize the opposite boundaries of the slab.  When the quantum states are outside  of the Dirac cone, they become bulk states, that is to say, the surface states only occur on the Dirac cone area of the surface Brillouin zone.

In order to investigate the pseudospin texture of the surface states, we define the pseudospin operators as $s_x= \frac{1}{2}\sigma_y\otimes \tau_x$, $s_y= \frac{1}{2}\sigma_x\otimes \tau_x$, and $s_z=-\frac{1}{2}\sigma_z\otimes I$ with the commutation relations
$[s_i, s_j]=i\epsilon_{ijk}s_k$.
For the surface states, the pseudospins form a vortex or antivortex pseudospin texture on the surface Brillouin zone with a number number $+1$ or $-1$, respectively. For the case in the parameter range (ii), Fig.\ref{surface1}(c)  shows the pseudospin texture of the average pseudospin components $(\langle s_x\rangle, \langle s_y\rangle)$ on the surface Brillouin zone and Fig.\ref{surface1}(d) shows that  the average pseudospin component $\langle s_z\rangle$ is  vanishing  for the Dirac surface states.
That is to say,  there is a singularity at the surface Dirac points $(-\pi/2, \pi/2)$ for the pseudospin texture. Integrating the Berry connection along  a circle enclosed the Dirac point for a valence band surface state, one can obtain a $\pi$ Berry phase, which can be used to define a winding number $-1$, so the pseudospin form an anti-vortex structure.  For the case in parameter range (iv), Fig.\ref{surface2}(c) and Fig.\ref{surface2}(d) show the pseudospin texture of the average pseudospin components $(\langle s_x\rangle, \langle s_y\rangle)$ on the surface Brillouin zone and the average pseudospin component $\langle s_z\rangle$, respectively. Similarly, the only the in-plane pseudospin components $(\langle s_x\rangle, \langle s_y\rangle)$  exist  and  they form a vortex.   The different point is that the vortex has a opposite winding number $+1$, compared with the case in the parameter range (ii).
Beside the parameter ranges (ii) and (iv), the parameter ranges (vi) and (viii) also support  the existence of $Z_2$ topological insulators. The  Dirac points have the same position and the surface states  have the same pseudospin texture on the surface Brillouin zone for the parameter ranges (ii) and (vi), and for the parameter ranges (iv) and (viii), respectively.

\section{The effects of the hidden-symmetry-breaking perturbations}
\begin{figure}[ht]

\includegraphics[width=0.45\columnwidth]{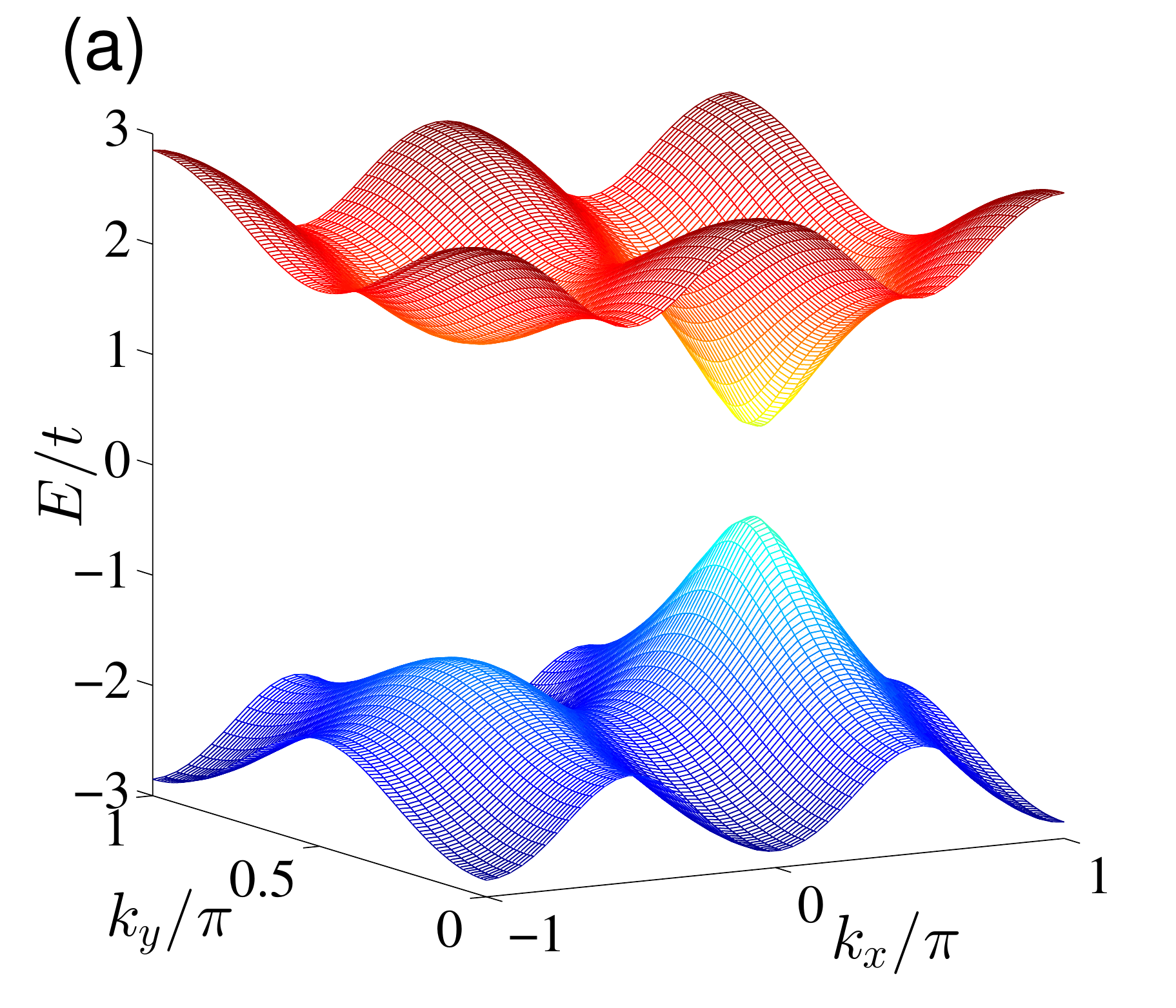}
 \includegraphics[width=0.45\columnwidth]{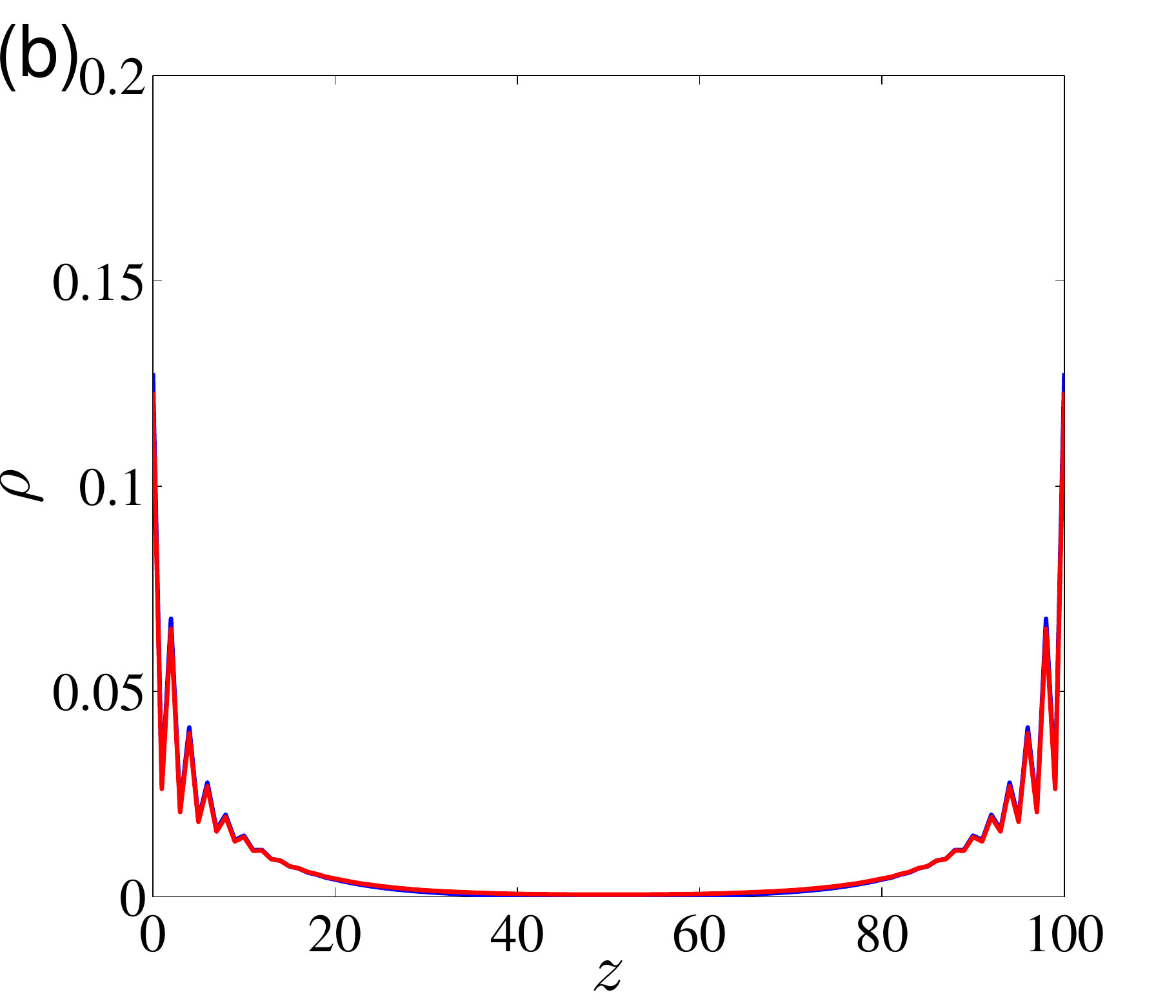}
 \includegraphics[width=0.45\columnwidth]{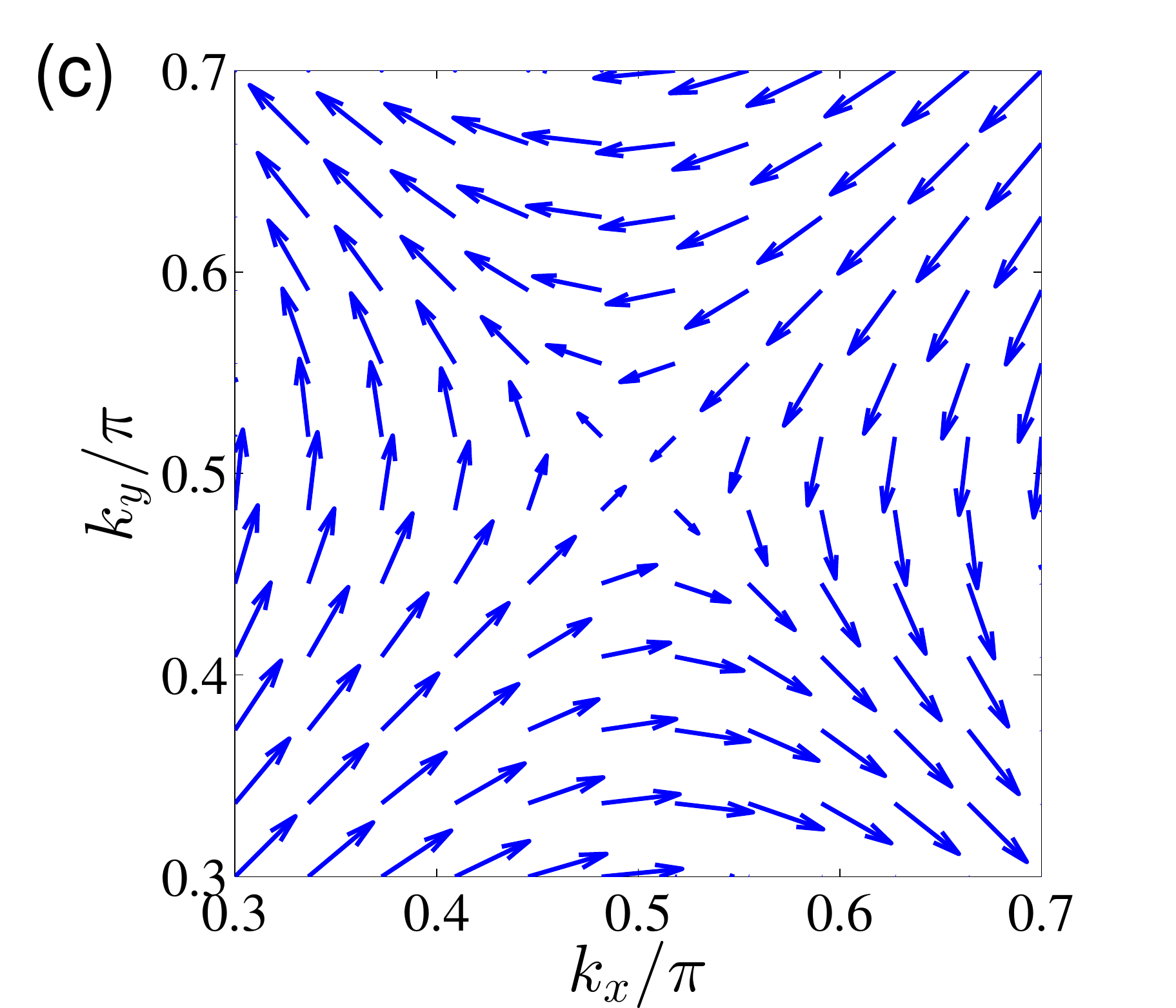}
   \includegraphics[width=0.45\columnwidth]{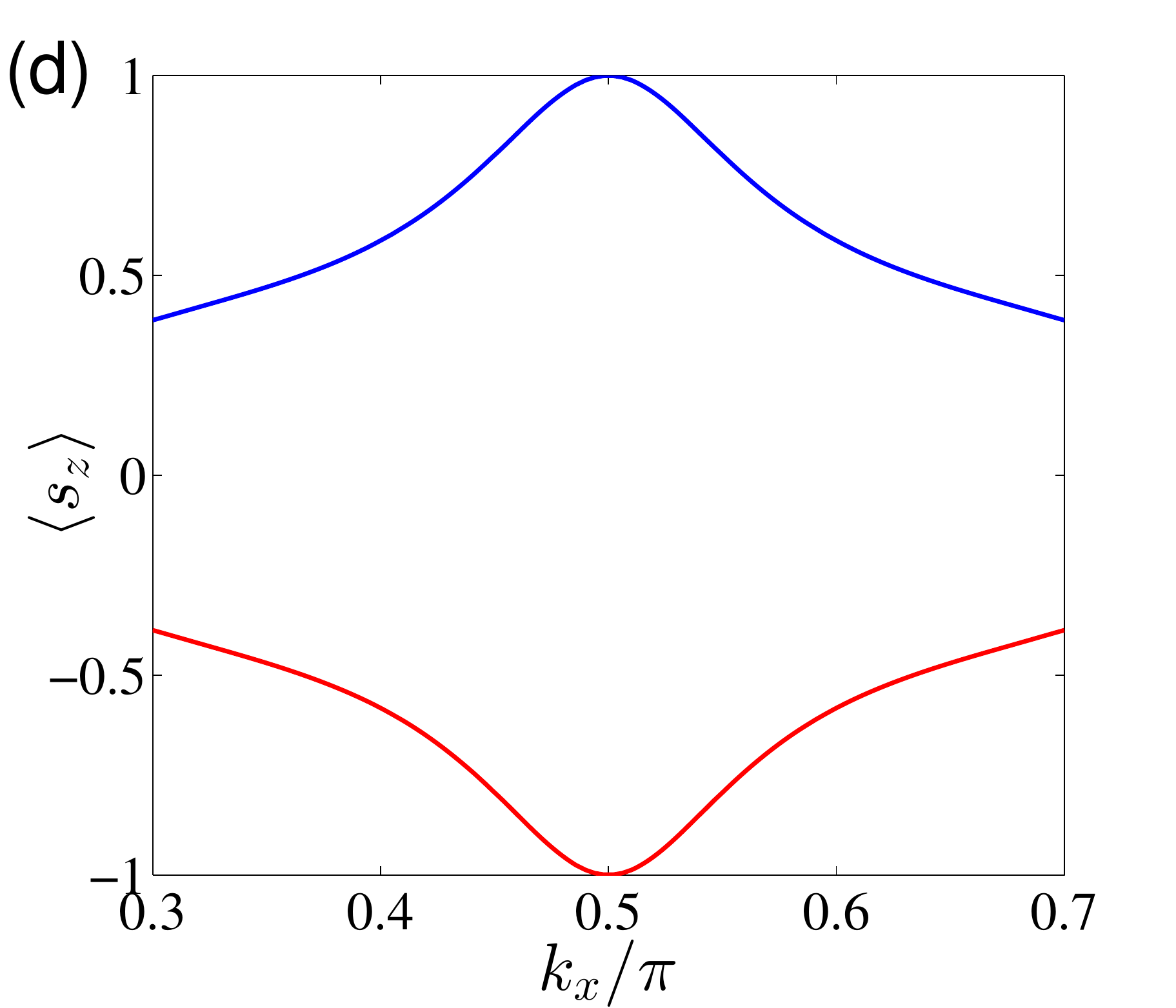}
\caption{The properties of the highest valence bands    of a slab with the hidden-symmetry-breaking perturbations for the case $t_1=0.3t$, $\lambda=t$, and $\mu=0.8t$.  (a) The dispersion relation;  (b)   the probability density of the states with the wave vector $(k_x,k_y)=(0.6\pi,0.5\pi)$, where the  lines correspond to the two quantum states are entirely overlapped; (c) the pseudospin textures of the average pseudospin components $(\langle s_x\rangle, \langle s_y\rangle)$ on the $k_x$-$k_y$ plane, which are identical and entirely overlapped  for two degenerate bands; (d) the profile of the average pseudospin component  $\langle s_z\rangle $ along the $k_y=0.5\pi$ line, where the red and blue lines correspond to the two degenerate states, respectively.   }\label{surfacedop1}
\end{figure}

\begin{figure}[ht]
\includegraphics[width=0.45\columnwidth]{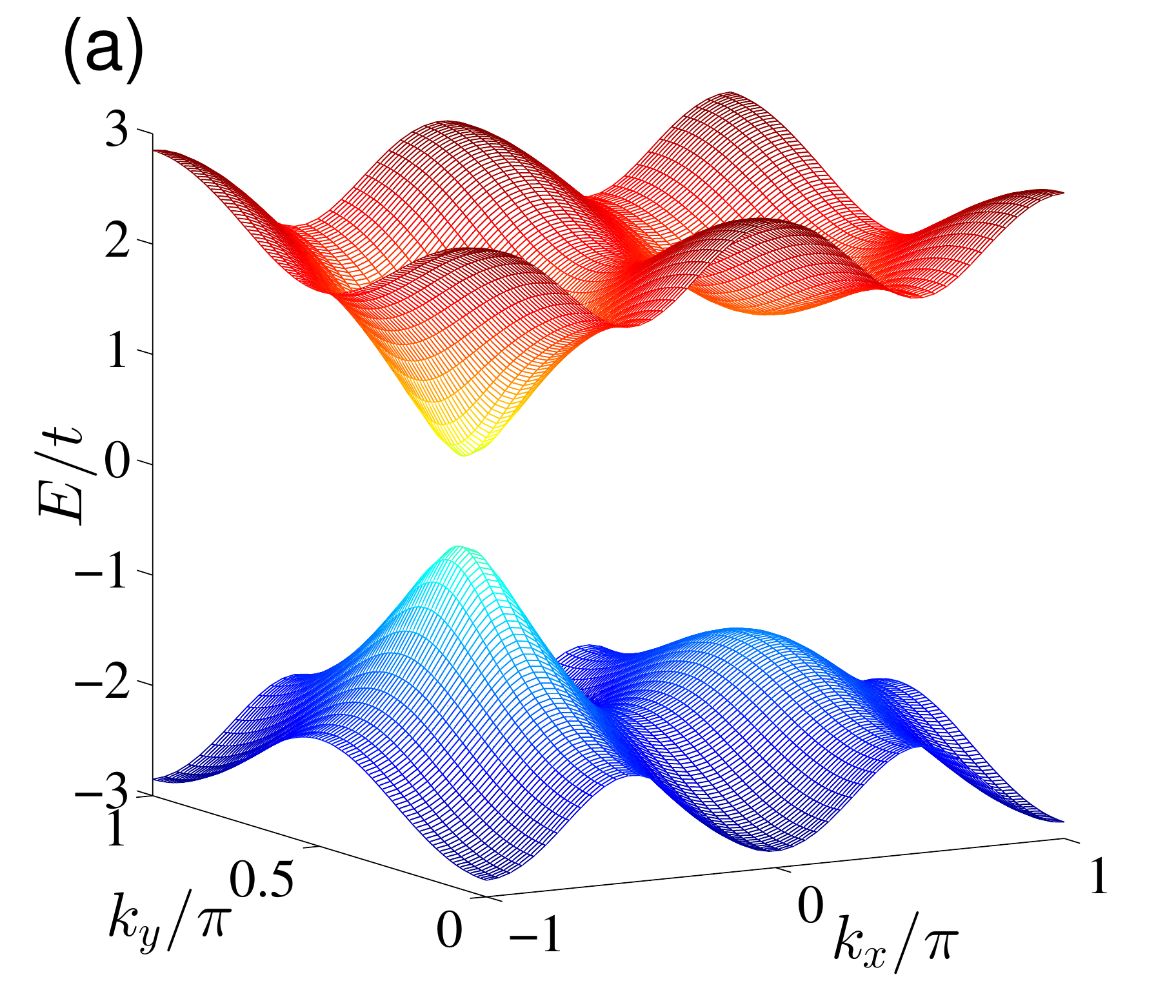}
\includegraphics[width=0.45\columnwidth]{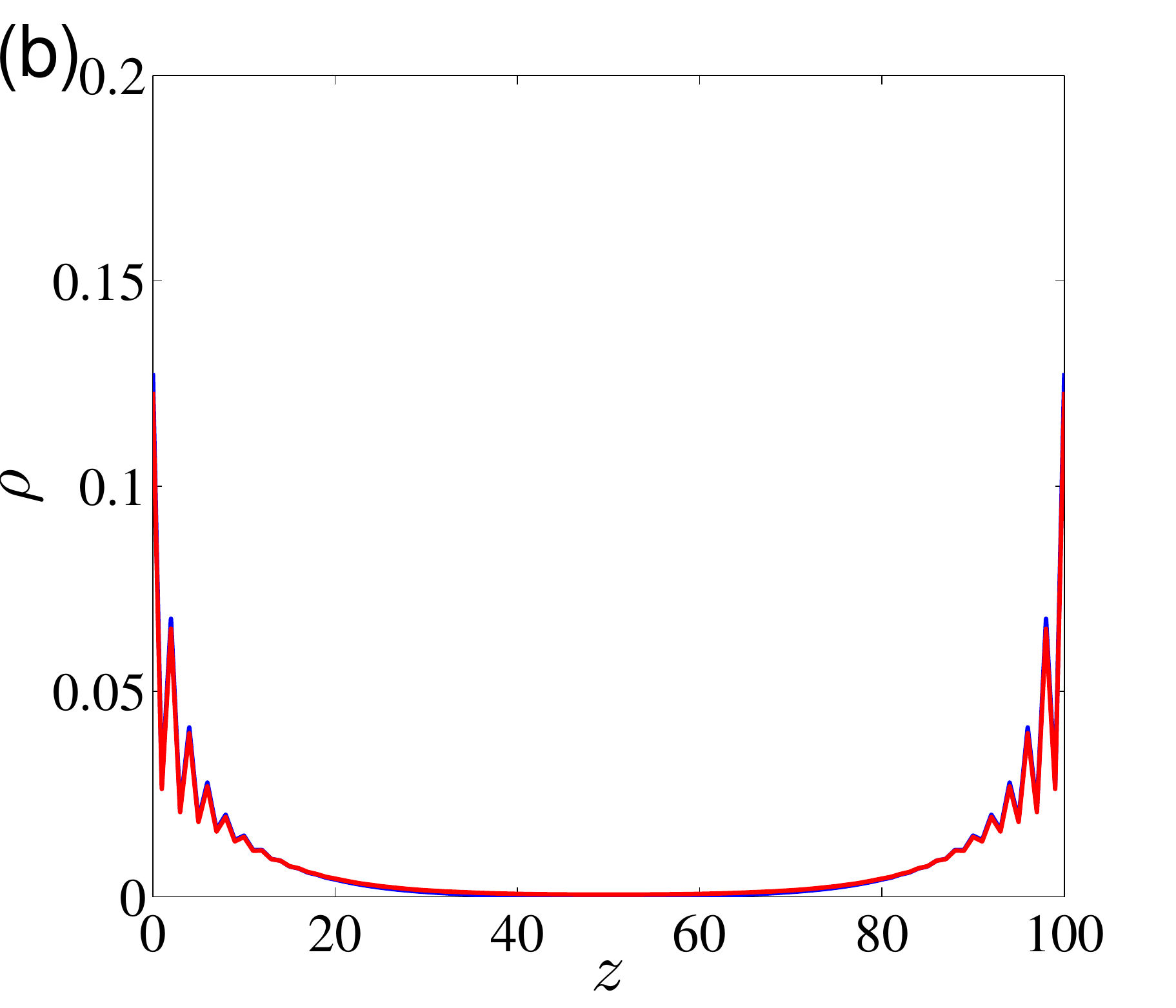}
 \includegraphics[width=0.45\columnwidth]{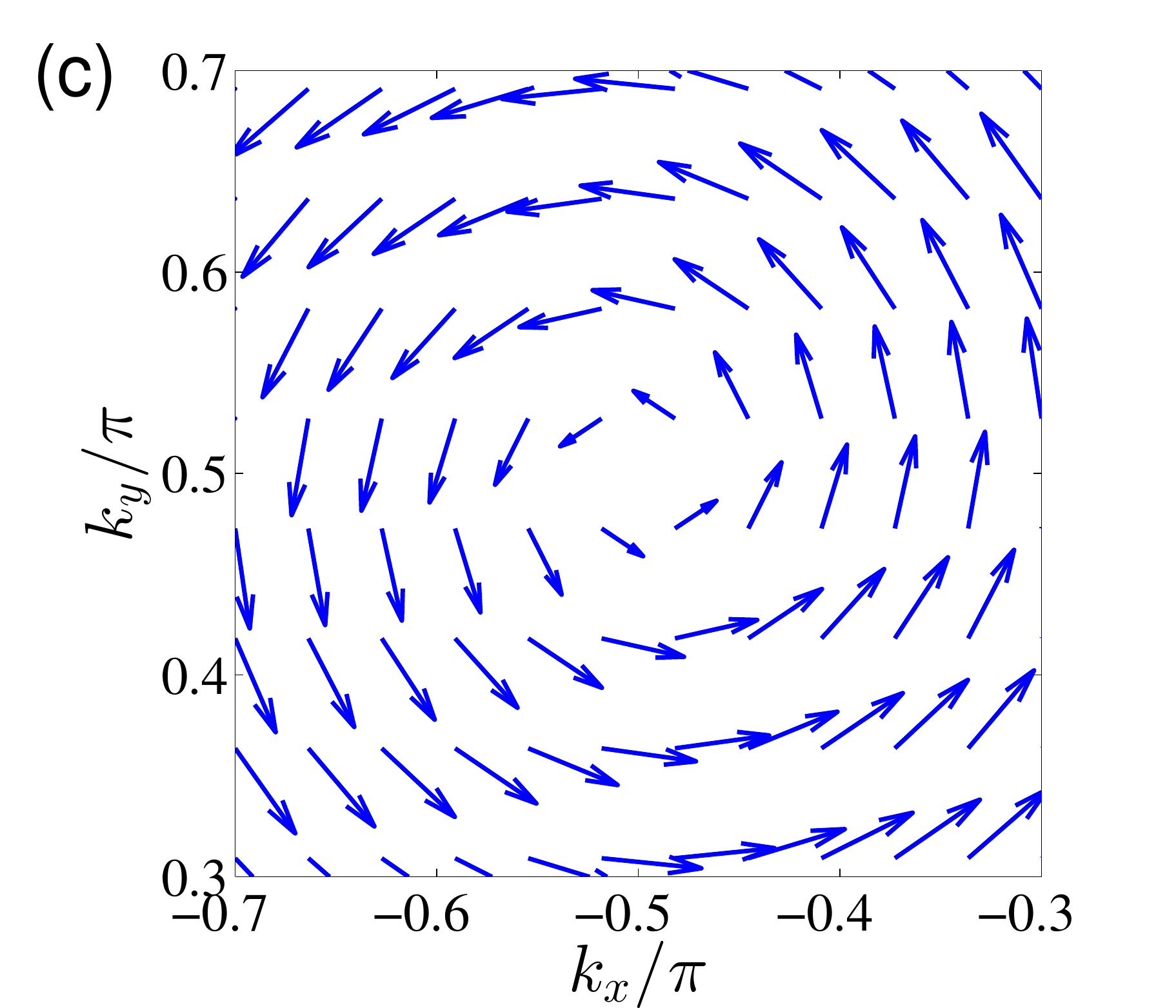}
  \includegraphics[width=0.45\columnwidth]{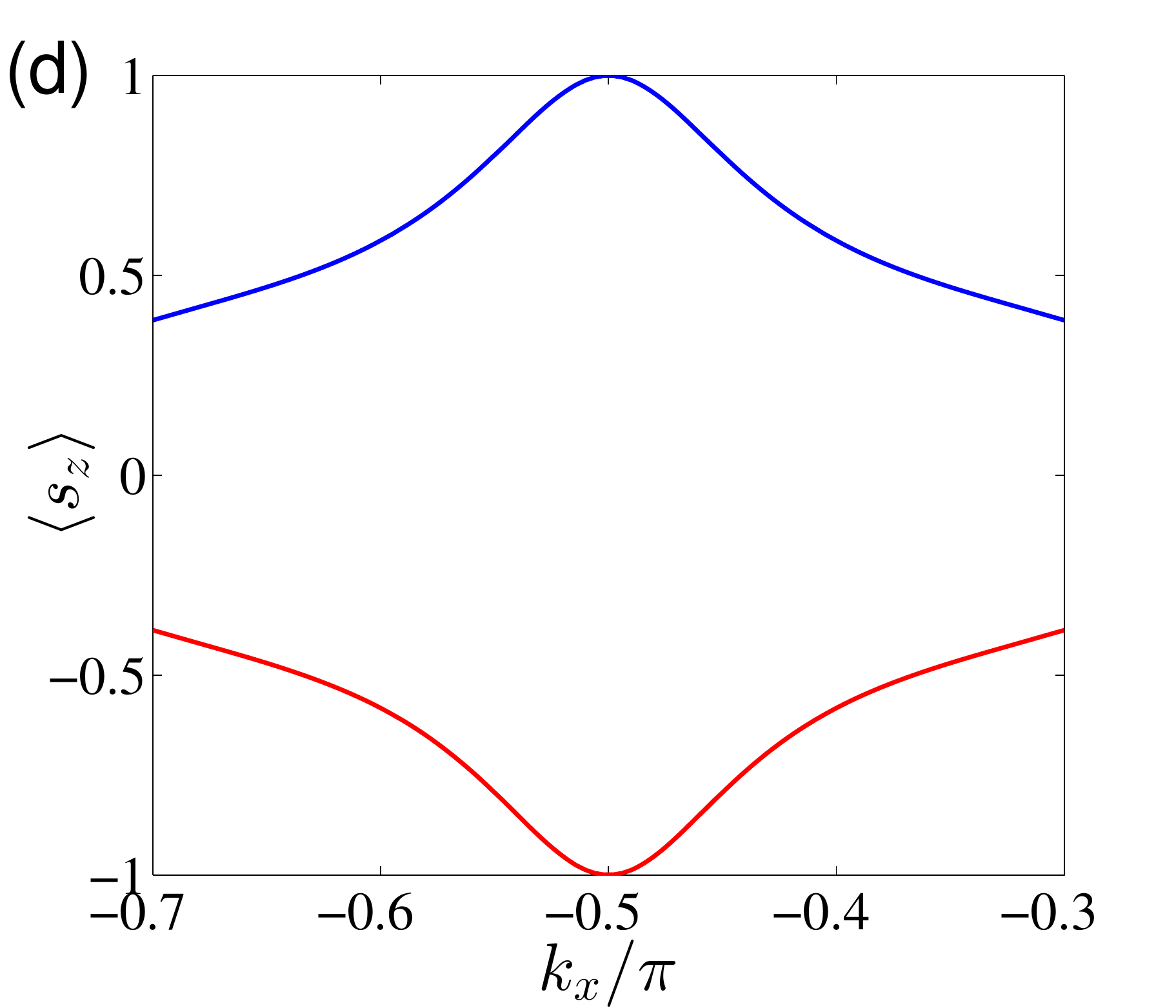}
\caption{ The properties of the highest valence bands    of a slab with the hidden-symmetry-breaking perturbations for the case $t_1=0.3t$, $\lambda=-t$, and $\mu=0.8t$. (a) The dispersion relation;  (b)   the probability density of the states   with the wave vector $(k_x,k_y)=(-0.6\pi,0.5\pi)$, where the  lines correspond to the two quantum states are entirely overlapped; (c) the pseudospin texture of the average pseudospin components $(\langle s_x\rangle, \langle s_y\rangle)$ on the $k_x$-$k_y$ plane, which are identical and entirely overlapped  for two degenerate bands; (d) the profile of the average  pseudospin component  $\langle s_z\rangle $ along the $k_y=0.5\pi$ line, where the red and blue lines correspond to the two degenerate states, respectively  }\label{surfacedop2}
\end{figure}

Since the $Z_2$ topological insulator and the surface states are protected by the hidden symmetry $\Upsilon$, the surface states should be gapped if some perturbations breaking the hidden symmetry $\Upsilon$ are added on the boundaries of the lattice. In order to verify the protection by hidden symmetry $\Upsilon$, we add  the hidden-symmetry-breaking perturbation terms  on the two open boundaries of a slab  and investigate the effects of these terms. We assume that the hidden-symmetry-breaking perturbations on the boundaries has the form,
\begin{eqnarray}
H_p=\mu\sum_{i\in SA} a^\dag_i a_i-\mu\sum_{j\in SB}b^\dag_jb_j
\end{eqnarray}
where $\mu$ is  the magnitude of the perturbations; $SA$ and $SB$ denote the boundary surfaces of sublattice $A$ and $B$.    We calculate the dispersion relations, probability densities, and pseudospin textures  for $t_1=0.3t$, $\lambda=t$, $\mu=0.8t$ and  $t_1=0.3t$, $\lambda=-t$, $\mu=0.8t$, which are shown in Fig.\ref{surfacedop1} and Fig.\ref{surfacedop2}, respectively.   Fig.\ref{surfacedop1}(a) and  Fig.\ref{surfacedop2}(a) show the highest   valence   bands and the lowest   conduction  bands, which correspond to the surface states in the case without hidden-symmetry-breaking perturbations. It is found that the hidden-symmetry-breaking perturbations open a gap between the highest valence and lowest conduction energy bands and the Dirac cones in the surface Brillouin zone disappear.
Fig.\ref{surfacedop1}(b) and Fig.\ref{surfacedop2}(b) show the probability density profiles of  the highest valence   states with the wave vectors $(0.6\pi, 0.5\pi)$ and  $(-0.6\pi, 0.5\pi)$, respectively, from which it is found that the mixing between the surface states on the two opposite boundaries happens due to the existence of the hidden-symmetry-breaking perturbations.  Fig.\ref{surfacedop1}(c) and Fig.\ref{surfacedop2}(c) show  the pseudospin textures of the highest valence states, which seem to  manifest similar vortex   pseudospin textures as the case without hidden-symmetry-breaking perturbations. In fact, for the case with hidden-symmetry-breaking perturbations,    the out-of-plane pseudospin component $\langle s_z\rangle$ appears as shown in Fig.\ref{surfacedop1}(d) and Fig.\ref{surfacedop2}(d), so the singularity of pseudospin texture disappears and the pseudospin textures are not vortices, which are consistent with the disappearing of the Dirac cones. The above characters confirm that the $Z_2$ topological insulator and the surface states are protected by the hidden  symmetry $\Upsilon$.

\section{Experimental techniques for realization and detection of the   $Z_2$ topological insulators with ultracold atoms in optical lattices }

During recent years, ultracold atoms in optical lattices have  become a platform to simulate the exotic physics in condensed matters, especially some of which are difficult to realize in real solid materials\cite{Jaksch05anp,Bloch08rmp,Lewenstein07adp}. A lot of   experimental
techniques  have been developed to construct various optical lattices, such as laser-assisting tunneling\cite{Aidelsburger11prl,Aidelsburger13prl,Miyake13prl,WuZ16sci}, shaking optical lattice\cite{Struck12prl,Struck13np}.  Due to the advantage of ultracold atoms in optical lattices, they
have often employed  to explore
the physics of topological phases\cite{Goldman16np,Goldman09prl,Goldman10prl,LiuGC10pra,Kennedy13prl,Bermudez10prl,Beri11prl}.
All of the techniques provide a foundation to design a model to realize a three-dimensional hidden-symmetry-protected $Z_2$ topological insulator in optical lattices.

In order to realize the model of Eqs.(\ref{tbh1}), (\ref{tbh2}), and (\ref{tbh3}), we   select two hyperspin states of cold atoms
  $^6$Li or $^{40}$K  to be trapped in an cubic optical
lattice formed by three pairs of lasers. These two hyperspin states can be regarded as the basis of the color space. The hyperspin-switching hopping can be induced by fine-designed Raman laser fields\cite{Aidelsburger11prl,Aidelsburger13prl,Miyake13prl,WuZ16sci}. The accompanying
phases of hopping can be realized by tuning the directions and frequencies of assistant  lasers\cite{ChenW17prb}.

  There also many techniques to detect the topological properties  in optical lattices. The atomic interferometry is a very useful technique to measure a relative phase. Based on this technique, a direct measurement of the Zak phase in topological Bloch bands was performed\cite{Atala13np} and an Aharonov-Bohm interferometer was constructed for determining Bloch band topology\cite{Duca15sci}. The concrete schemes were designed to measure Chern number\cite{Abanin13prl} and   $Z_2$ topological invariant \cite{Grusdt14pra} with atomic interferometry. Very recently, Bloch state tomography was developed to detect Berry curvature and topological invariants, including single- and multiband Chern and $Z_2$ numbers\cite{LiT16sci}. Another detecting technique is  Bragg spectroscopy\cite{Stamper-Kurn99prl}, which can be employed to probe the dispersion relation of optical lattice. A scheme to detect the edge states based on Bragg  scattering  was proposed\cite{Goldman12prl}. The edge states can also be detected by direct imaging method\cite{Goldman13pnas}.
  The high-resolution
uorescence imaging can probe   optical lattices at single-site level\cite{Bakr10sci,Sherson10nat},   so it can be employed to detect edge or surface states by measuring  populations of atoms.  Based on the above detecting techniques, it is feasible that the $Z_2$ topological invariant and the surface states in the hidden-symmetry-protected $Z_2$ topological insulators are measured and detected.

\section{Conclusion}

In summary, we have studied a tight-binding model in a cubic lattice that preserves a hidden  symmetry, which has a composite antiunitary operator consisting of fractional translation, complex conjugation, sublattice
exchange, and local gauge transformation. Based on the hidden symmetry, we defined a $Z_2$ topological invariant which classifies the insulator phases of  the lattice. In some parameter ranges, the lattice supports a non-trivial topological insulator protected by the hidden symmetry. For the hidden-symmetry-protected topological insulator,  the surface states localized on one of open boundaries and have a single Dirac cone band structure on the surface Brillouin zone. We also defined   pseudospin operators and find that the surface states have only   in-plane components, which form a vortex or antivortex pseudospin texture having a winding number $\pm 1$, respectively.  When additional hidden-symmetry-breaking perturbations on the open boundaries of a slab geometry are added,  an energy gap opens between the highest valance  and   lowest conduction bands and the surface states on the two opposite boundaries mix, even turn into bulk states when the perturbations are strong enough. Furthermore, the out-of-plane pseudospin component appears and the singularity of pseudospin texture vanishes. The results of the case with additional hidden-symmetry-breaking perturbations demonstrate that the $Z_2$ topological insulator and surface states are protected by the hidden symmetry $\Upsilon$.

\begin{acknowledgments}
   This work was supported by the National Natural
Science Foundation of China under Grants No. 11274061 (J.M.H.) and No. 11504171
(W.C.).  W.C. was also supported
by the Natural Science Foundation of Jiangsu Province in
China under Grant No. BK20150734.
\end{acknowledgments}

\end{document}